\documentclass[12pt, draftclsnofoot, onecolumn]{IEEEtran}
\usepackage{amsmath}
\usepackage{amssymb}
\usepackage{amsthm}
\usepackage{cite}
\usepackage{color}
\usepackage{xcolor}

\usepackage{caption}
\usepackage{epsfig,latexsym}
\usepackage{float}
\usepackage{fancyhdr}
\usepackage{graphicx}
\usepackage{indentfirst}
\usepackage{mdwmath}
\usepackage{mdwtab}
\usepackage{subfigure}
\usepackage{setspace}
\usepackage{times}
\usepackage{url}

\newtheorem{remark}{Remark}
\newtheorem{theorem}{Theorem}

\newtheorem{lemma}{Lemma}

\newtheorem{corollary}{Corollary}

\begin{document}

\title{Exploiting NOMA for UAV Communications in Large-Scale Cellular Networks}
\author{Tianwei Hou,~\IEEEmembership{Student Member,~IEEE,}
        Yuanwei Liu,~\IEEEmembership{Senior Member,~IEEE,}\\
        Zhengyu Song,
        Xin Sun,
        Yue Chen,~\IEEEmembership{Senior Member,~IEEE,}

\thanks{This work is supported by the Fundamental Research Funds for the Central Universities under Grant 2019YJS008. Part of this work was submitted to the IEEE Vehicular Technology Conference, Hawaii, USA, Fall 2019~\cite{multi_VTC2019}.}
\thanks{T. Hou, Z. Song and X. Sun are with the School of Electronic and Information Engineering, Beijing Jiaotong University, Beijing 100044, China (email: 16111019@bjtu.edu.cn, xsun@bjtu.edu.cn, songzy@bjtu.edu.cn).}
\thanks{Y. Liu and Yue Chen are with School of Electronic Engineering and Computer Science, Queen Mary University of London, London E1 4NS, U.K. (e-mail: yuanwei.liu@qmul.ac.uk, yue.chen@qmul.ac.uk).}
}

\maketitle
\vspace{-1cm}

\begin{abstract}
This paper advocates a pair of strategies in non-orthogonal multiple access (NOMA) in unmanned aerial vehicles (UAVs) communications, where multiple UAVs play as new aerial communications platforms for serving terrestrial NOMA users. A new multiple UAVs framework with invoking stochastic geometry technique is proposed, in which a pair of practical strategies are considered: 1) the UAV-centric strategy for offloading actions and 2) the user-centric strategy for providing emergency communications. In order to provide practical insights for the proposed NOMA assisted UAV framework, an imperfect successive interference cancelation (ipSIC) scenario is taken into account. For both UAV-centric strategy and user-centric strategy, we derive new exact expressions for the coverage probability. We also derive new analytical results for orthogonal multiple access (OMA) for providing a benchmark scheme.
The derived analytical results in both user-centric strategy and UAV-centric strategy explicitly indicate that the ipSIC coefficient is a dominant component in terms of coverage probability.
Numerical results are provided to confirm that i) for both user-centric strategy and UAV-centric strategy, NOMA assisted UAV cellular networks is capable of outperforming OMA by setting power allocation factors and targeted rate properly; and ii) the coverage probability of NOMA assisted UAV cellular framework is affected to a large extent by ipSIC coefficient, target rates and power allocations factors of paired NOMA users.
\end{abstract}

\begin{IEEEkeywords}
Non-orthogonal multiple access, stochastic geometry, unmanned aerial vehicles.
\end{IEEEkeywords}

\section{Introduction}

\IEEEPARstart{I}{n} the past decades, much research effort has been directed towards developing remotely operated unmanned aerial vehicles (UAVs), which stand as a potential candidate of aerial base station (BS) to provide access services to wireless devices located on the ground~\cite{UAV_zeng} or in the sky~\cite{Saad_D2D_UAV}. UAV communications are also an effective approach to provide connectivity during temporary events and after disasters in the remote areas that lack cellular infrastructure~\cite{UAV_zeng}.
As compared to conventional terrestrial communications, one distinct feature of UAV communication is that the existence of line-of-sight (LoS) is capable of offering stronger small-scale fading between UAVs and ground users because of the high altitude of UAVs, which brings both opportunities and challenges in the design of UAV cellular networks~\cite{3GPP_36.777}.
Due to the limited energy resources on board of a UAV, achieving higher spectrum efficiency and energy efficiency is of paramount importance to reap maximum benefits from UAV based communication networks.

To exploit both the spectrum efficiency and energy efficiency in the next generation wireless networks and beyond, especially in the UAV communication networks, non-orthogonal multiple access (NOMA) is considered to be a promising technique~\cite{NOMA_mag_Ding,wireless_sparse}.
More specifically, in contrast to the conventional OMA techniques, NOMA is capable of exploiting the available resources more efficiently by opportunistically capitalizing on the users’ specific channel conditions on both single cell networks and cellular networks~\cite{PairingDING2016,Massive_NOMA_Cellular_IoT}, and it is capable of serving multiple users at different quality-of-service (QoS) requirements in the same resource block~\cite{NOMA_5G_beyond_Liu,Resource_allo_Islam,Islam_NOMA_survey}. To be more clear, NOMA technique sends the composite signal to multiple users simultaneously by power domain multiplexing within the same frequency, time and code block. The basic principles of NOMA techniques rely on the employment of superposition coding (SC) at the transmitter and successive interference cancelation (SIC) techniques at the receiver~\cite{NOMA_mag_Ding,heterNOMA_Qin}, and hence multiple accessed users can be realized in the power domain via different power levels for users in the same resource block\footnote{In this paper, we use “NOMA” to refer to “power-domain NOMA” for simplicity.}. Therefore, UAV networks can serve multiple users simultaneously by utilizing NOMA techniques for enhancing the achievable spectrum efficiency and energy efficiency.

\subsection{Prior Work and Motivation}

Regarding the literature of UAV networks, early research contributions have studied the performance of single UAV or multiple UAVs networks. Mozaffari {\em et al.}~\cite{Saad_D2D_UAV} proposed a UAV assisted underlaid D2D network with LoS probability, which depends on the height of the UAV, the horizontal distance between the UAV and users, the carrier frequency and type of environment. In the case that LoS exists, a fixed LoS coefficient, e.g., an extra 20dB attenuation, is the dominant component of small-scale fading channels.
Note that the proposed model in~\cite{3GPP_36.777,Saad_D2D_UAV} is a practical model for implementation. For mathematically tractable, the distinctive channel characteristics for UAV networks were investigated in~\cite{UAV_Channel}, where different types of small-scale fading channels, i.e., Loo model, Rayleigh model, Nakagami-$m$ model, Rician model and Werbull model, were summarized to demonstrate the channel propagation of UAV networks.
The air-to-air channel characterization in~\cite{A2A_UAV_Rice}, studied the influence of the altitude–-dependent Rician K factor. This work indicated that the impact of the ground reflected multi-path fading reduces with increasing UAV altitude. Jiang~{\em et al.}~\cite{Rayleigh_UAV} proposed a UAV assisted ground-to-air network, where Rician channels are used for evaluating strong LoS components between UAV and ground users. It is also worth noting that Rayleigh fading channel, which is a well-known model in scattering environment, can be also used to model the UAV channel characteristics in the case of large elevation angles in the mixed–-urban environment. Chetlur {\em et al.}~\cite{UAV_finite_downlink} proposed a downlink UAV network over Nakagami-$m$ fading channels, where UAVs are distributed in a finite 3-D network. An uniform binomial point process was invoked to model the proposed network. Generally speaking, Nakagami-$m$ distribution and Rician distribution are used to approximate the fluctuations in the fading channel with LoS propagations. It is also worth noting that the fading parameter of Nakagami-$m$ fading $m=\frac{(K+1)^2}{2K+1}$, the distribution of Nakagami-$m$ is approximately Rician fading with parameter $K$~\cite[eq. (3.38)]{wireless_communication_goldsmith}. Zhang~{\em et al.}~\cite{UAV_Trajectory_shuowen} proposed two possible paradigms for UAV assisted cellular communications, namely, cellular-enabled UAV communication and UAV-assisted cellular communication. The trajectory of the UAV was optimized under connectivity-constrained. Lyu~{\em et al.}~\cite{Lyu_UAV_hotspots} proposed a UAV assisted cellular hotspot scenario, where UAV flies cyclically along the cell edge for offloading actions.
In order to improve the spectrum efficiency and energy efficiency of UAV communications, new research on UAV under emerging next generation network architectures is needed.

Recently, the use of NOMA in wireless communication has attracted great interest in single cell or cellular networks~\cite{Islam_NOMA_survey,Dai_NOMA_survey,Shin_NOMA_cellular}. Ding {\em et al.}~\cite{Randomly_ding} evaluated the performance of NOMA enhanced single cell networks with randomly deployed users, where order statistics and stochastic geometry tools were invoked to evaluate the performance of paired NOMA users. Some application scenarios of NOMA have been investigated in the previous literature. More particularly, Liu {\em et al.}~\cite{Liu_Coop_NOMA_SWIPT} proposed an innovative model of cooperative NOMA with simultaneous wireless information and power transfer (SWIPT), where a NOMA cluster consists of two NOMA users, one that is located in a small disk and the other is in a ring with a larger external radius.
Ding {\em et al.}~\cite{PairingDING2016} evaluated the performance of NOMA with fixed power allocation (F-NOMA) and cognitive radio inspired NOMA (CR-NOMA), and the user pairing strategies were carefully discussed. The analytical results show that it is more preferable to pair users whose channel gains are more distinctive to improve the diversity order in F-NOMA, whereas CR-NOMA prefers to pair the users with the best channel conditions.
Recently, an imperfect SIC scenario has attracted great interest. Due to the fact that SIC techniques are deployed at the receivers, the residues of the multiplexed signal detected by SIC technique cannot be ignored~\cite{Yue_ISIC_2018}. Once an error occurs for carrying out SIC at the user with better channel gain, the NOMA systems will suffer from the residual interference signal. Hence it is significant to examine the detrimental impacts of imperfect SIC for NOMA system. Hou {\em et al.}~\cite{Nakagami_Hou} evaluated the outage performance of NOMA downlink transmission in both LoS and NLoS scenarios.
A potential future research direction for NOMA, called Rate-Splitting multiple access, has been proposed by Mao {\em et al.}~\cite{Clerckx_RSMA}. The analytical results in~\cite{Clerckx_RSMA} demonstrated that RSMA can outperform SDMA and NOMA in the multi-antenna system and comes with a lower complexity than NOMA. RSMA assisted multi-cell networks and multi-antenna assisted RSMA were also proposed in~\cite{Clerckx_RSMA_MIMO}. The results derived concluded that RSMA can provide rate, robustness and QoS enhancements over SDMA and NOMA. With the goal of enhancing the physical layer security of NOMA networks, Liu \emph{et al.}~\cite{Liu_physical_scurity_NOMA} proposed a NOMA assisted physical layer security framework in large-scale networks, where both single antenna and multiple antenna aided transmission scenarios were considered.

In UAV-enabled wireless communications, the total UAV energy is limited, which includes propulsion energy and communication related energy~\cite{energy_consumption_UAV}. Therefore, integrating UAVs and NOMA into cellular networks is considered to be a promising technique to significantly enhance the performance of terrestrial users in the next generation wireless system and beyond, where the energy efficiency and spectrum efficiency can be greatly enhanced in downlink transmission to minimize communication related energy~\cite{Cellular-UAV_mag}. A general introduction of UAV communications has been proposed by Liu {\em et al.}~\cite{UAV_general_Liu}. Three case studies, i.e., performance evaluation, joint trajectory design, and machine learning assisted UAV deployment~\cite{Liuxiao_Trajectory_multi-UAV}, were carried out in order to better understand NOMA enabled UAV networks. Some challenges were concluded for future research directions. Zhao~{\em et al.}~\cite{UAV_NOMA_Trajectory} proposed a UAV-assisted NOMA network, where UAV and BS are cooperated to provide access services to ground users simultaneously. The trajectory of UAV and precoding matrix of BS were jointly optimized. Nguyen~{\em et al.}~\cite{UAV_relay} proposed a cooperative multi-UAV network, where a fixed number of UAVs are used as flying relays in wireless backhaul networks, and the small-scale fading follows Rician distributions. Hou {\em et al.}~\cite{Hou_Single_UAV} proposed a multiple-input multiple-output (MIMO)-NOMA assisted UAV network, where the closed-form expressions of outage performance and ergodic rate were evaluated in the downlink scenario. A NOMA assisted uplink scenario of UAV assisted cellular communication was proposed by Mei~{\em et al.}~\cite{Uplink_NOMA_UAV}, where two special cases, i.e., egoistic and altruistic transmission strategies of the UAV, were considered to derive the optimized solutions. Liu~{\em et al.}~\cite{UAV_multibeam_liangliu} proposed a MIMO-NOMA assisted UAV network for uplink transmission, where the cellular-connected UAV communication with air-to-ground interference was investigated by utilizing multi-beam techniques. Han {\em et al.}~\cite{Han_millimeter_UAV} proposed a UAV assisted millimeter-wave air-to-everything networks, where aerial access points provide access services to users located on the ground, air, and tower. The buildings were modeled as a Boolean line-segment process with the fixed height.

The previous contributions~\cite{Satellite_UAV_network,Hou_Single_UAV,UAV_general_Liu,Uplink_NOMA_UAV,UAV_multibeam_liangliu,UAV_relay,Han_millimeter_UAV} mainly consider NOMA in single UAV cell or NOMA assisted uplink transmission, and thus do not account for NOMA assisted downlink transmission in UAV assisted cellular networks. The research contributions in terms of conducting on multi-UAV aided NOMA networks are still in their infancy, particularly with the focus of potential association strategies. NOMA enhanced UAV networks design poses three additional challenges: i) NOMA technology brings additional intra-cell interference from the connected UAV to the served users; ii) UAV communication requires different fading channels to evaluate the channel gain of LoS/NLoS propagation. iii) the user association policy requires to be reconsidered in NOMA assisted UAV networks.
In this article, aiming at tackling the aforementioned issues, by proposing two potential association strategies, namely UAV-centric strategy and user-centric strategy, for intelligently investigating the effect of NOMA assisted UAV network performance is desired.
The motivation of proposing two strategies is that the user-centric strategy is a promising solution for providing access services after disasters in the remote areas, where all of terrestrial users located in the Voronoi cell can be served by UAVs. On the contrary, the UAV-centric strategy can be perfectly deployed in the dense networks, i.e., concerts or football matches, to provide supplementary access services for offloading actions, where terrestrial users are located in a regular disc. Note that one other non-negligible difference between the two strategies is that user association is decided by individual user or UAV for the user-centric strategy or the UAV-centric strategy, respectively. Stochastic geometry tools are invoked to provide the mathematical paradigm to model the spatial randomness of both UAVs and users in UAV cellular networks. In contrast to the conventional terrestrial communication structure, where the locations of BSs are fixed, stochastic geometry is more suitable for analyzing the average performance of the mobility and flexibility of the UAV networks.

\subsection{Contributions}

In contract to most existing research contributions in context of UAV communications~\cite{Satellite_UAV_network,Hou_Single_UAV,UAV_general_Liu,Uplink_NOMA_UAV,UAV_multibeam_liangliu,UAV_relay,Han_millimeter_UAV}, we consider a multi-cell set-up in this paper. We propose two new NOMA assisted UAV cellular strategies, namely user-centric strategy and UAV-centric strategy.
Based on the proposed strategies, the primary theoretical contributions can be summarized as follows:

\begin{itemize}
  \item We develop two potential association strategies to address the impact of NOMA on the UAV communications, namely user-centric strategy and UAV-centric strategy, where stochastic geometry approaches are invoked to model the locations of both UAVs and users.
  \item For the user-centric strategy: we derive the exact analytical expressions of a typical user in the NOMA enhanced user-centric strategy in terms of coverage probability. Additionally, we derive the exact expressions in terms of coverage probability for the OMA assisted user-centric strategy. Our analytical results illustrate that the distance of the fixed user has effect on the coverage probability of the typical user. Furthermore, for the case of poor SIC quality, a hybrid NOMA/OMA assisted UAV framework may be a good solution.
  \item For the UAV-centric strategy: we derive the exact analytical expressions of paired NOMA users in the NOMA enhanced UAV-centric strategy in terms of coverage probability. The exact expressions in terms of coverage probability for the OMA case are derived. Our analytical results indicates that the UAV-centric strategy is more susceptible to ipSIC factor than the user-centric strategy.
  \item Simulation results confirm our analysis, and illustrate that by setting power allocation factors and targeted rate properly, NOMA assisted UAV cellular frameworks has superior performance over OMA assisted UAV cellular frameworks in terms of coverage probability, which demonstrates the benefits of the proposed strategies. Our analytical results also illustrate that the coverage probability can be greatly enhanced by LoS links.
\end{itemize}

\subsection{Organization}

The rest of the paper is organized as follows. In Section \uppercase\expandafter{\romannumeral2}, the NOMA assisted user-centric strategy is investigated for UAV cellular frameworks, where the UAV provides access services to all the users. In Section \uppercase\expandafter{\romannumeral3}, the NOMA assisted UAV-centric strategy is investigated, where the UAV only provides access services to the restricted areas. Our numerical results are demonstrated in Section~\uppercase\expandafter{\romannumeral4} for verifying our analysis, which is followed by the conclusion in Section \uppercase\expandafter{\romannumeral5}.

\section{User-centric Strategy for Emergency Communications}

We first focus our attention on a scenario, where all the terrestrial users are needed to be served equally for emergency communications, e.g., after disasters, in the remote areas or in the rural areas~\cite{UAV_emergency_disasters}. Motivated by this purpose, we propose the user-centric strategy for providing emergency access services to all the terrestrial users.

\begin{figure*}[t!]
\centering
\subfigure[Illustration of the system model.]{\label{system model}
\includegraphics[width =2.5in]{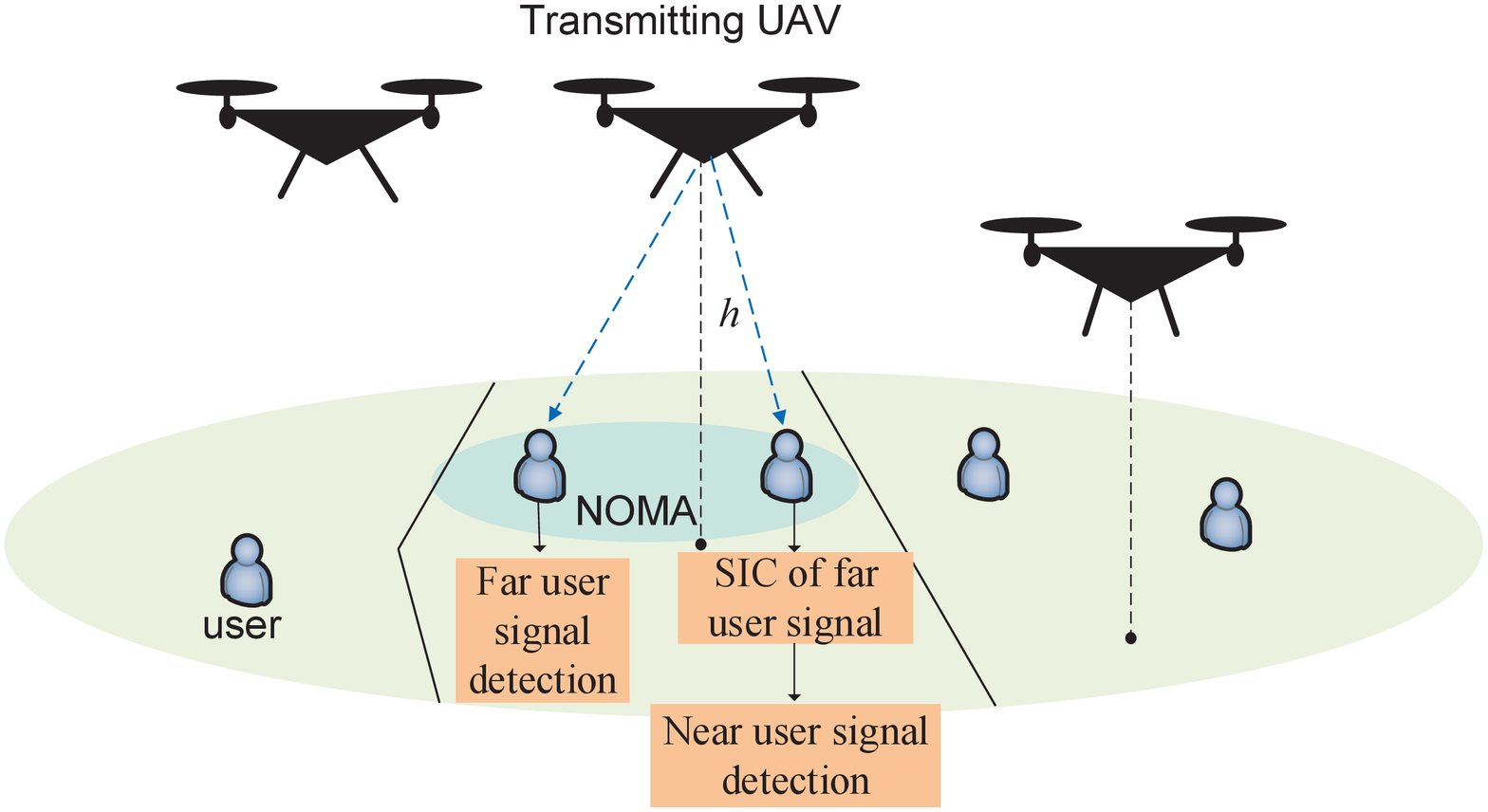}}
\subfigure[Top view of the user-centric strategy cellular networks.]{\label{user-centric example}
\includegraphics[width= 2.5in]{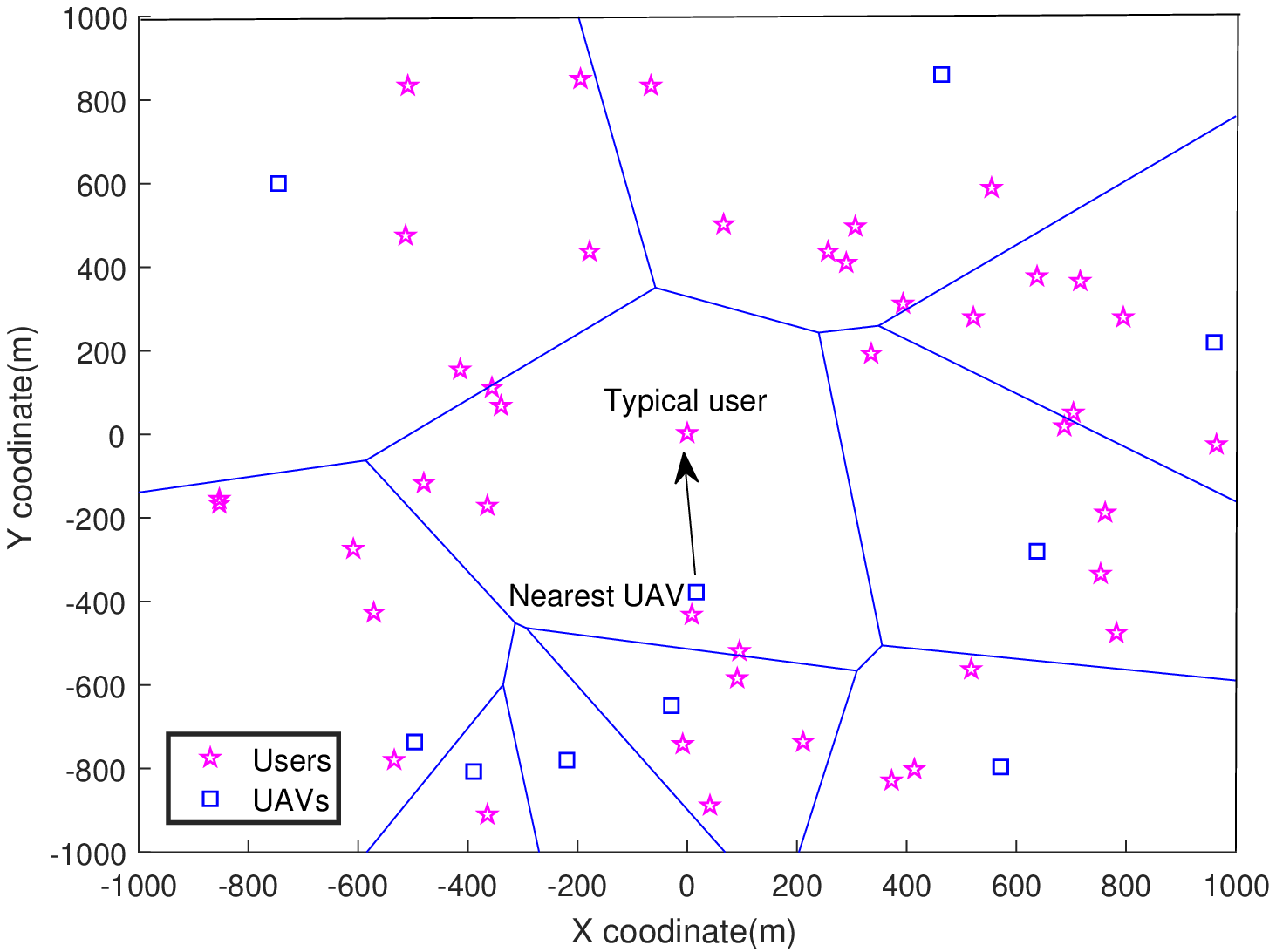}}
\caption{Illustration of NOMA assisted user-centric strategy model.}
\label{Example of User-Centric}
\vspace{-0.3in}
\end{figure*}

Focusing on downlink transmission scenarios, we consider the user-centric strategy as shown in Fig.~\ref{system model}. In this article, the UAV equipped with a single antenna communicates with multiple users equipped with a single antenna each.
In the user-centric strategy, the locations of terrestrial users are totally random for emergency services, and there are no further information for UAVs to properly organize their trajectory. In order to serve all the terrestrial users equally, multiple UAVs should be distributed uniformly, which conforms the definition of homogeneous poisson point process (HPPP). Thereby, the UAVs are distributed according to a HPPP $\Psi$ with density $\lambda$.
For the simplicity of theoretical analysis, as shown in Fig.~\ref{user-centric example}, an user is located at the original point in the user-centric strategy, which becomes the typical user. The user-centric strategy is a useful model for the large-scale networks, i.e., rural area, where users are uniformly located in the Voronoi cell according to a HPPP $\Phi_u$ with density $\lambda_u$. It is worth mentioning that in the case that the density of user $\lambda_u$ is low, the user-centric performs much better than the UAV-centric strategy.

Without loss of generality, we consider that there is one user, namely fixed user, is already connected to the UAV in the previous round of user association process\footnote{In practice, multiple users are connected to the transmitter (UAV) one by one.}. For simplicity, we assume that the horizontal distances between the fixed user and the connected UAV is $r_k$, which can be any arbitrary values, and the horizontal distance between the typical user and the connected UAV is random, denoted by $r$. In the user-centric strategy, we consider that two users, fixed user and typical user, are paired to perform NOMA technique, where paired NOMA users share the same frequency, time and code resource blocks.

\subsection{Channel Model}

Consider the use of a composite channel model with two parts, large-scale fading and small-scale fading. It is assumed that the horizontal distance $r$ and the height of the UAV $h$ are independent and identically distributed (i.i.d.). In this article, large-scale fading represents the path loss between the UAV and users\footnote{A log-normal distributed random variable for shadowing on both the desired and interference signals was considered in~\cite{no_shadowing}, which indicates that Lognormal shadowing on both the desired and interfering signals does not significantly affect the accuracy of numerical analysis. Thus, we neglect it in this article for simplicity.}.

In order to better illustrate the LoS propagation between the UAV and user, the small-scale fading is defined by Nakagami-\emph{m} fading, and the probability density function (PDF) can be expressed as
\begin{equation}\label{channel matrix,eq3}
{f}(x) = \frac{{m^m {x^{{m} - 1}}}}{{\Gamma ({m})}}{e^{ - {{mx}}}},
\end{equation}
where $m$ denotes the fading parameter, and $\Gamma ({m})$ denotes Gamma function. Note that $\Gamma ({m})=(m-1)!$ when $m$ is an integer. The serving distance between the connected UAV to the typical user can be written as
\begin{equation}\label{projective_distance,eq4}
{r_t} = \sqrt {{h^2} + {r^2}},
\end{equation}
where $r$ is the nearest horizontal distance allowed between a typical user and its connected UAV.

In order to avoid infinite received power, it is assumed that the height of the UAV is greater than 1m to simplify the analytical results.
Therefore, the large-scale fading can be expressed as
\begin{equation}\label{large-scale fading}
L_t={r_t^{ - \alpha }} ,
\end{equation}
where $\alpha$ denotes the path loss exponent between the typical user and the connected UAV. Thus, the received power from the associated UAV for the user at origin is given by
\begin{equation}\label{received user power}
{P_t} = {P_u}{L_t}{\left| {{{h_t}}} \right|^2},
\end{equation}
where $P_u$ denotes the transmit power of the UAV, and $h_{t}$ denotes the channel coefficients for the typical user and its associated UAV.

In downlink transmission, paired NOMA users also detect interference from neighboring UAVs. Therefore, the co-channel interference ${I}$ can be further expressed as follows:
\begin{equation}\label{interference,eq7}
{I} \buildrel \Delta \over = \sum\limits_{j \in \Psi, d_j>{r_{t}} } {\left| {{{g_j}}} \right|^2} {{P _u}d_{j}^{ - \alpha_I }}  ,
\end{equation}
where $d_{j}$ and ${\left| {{{g_j}}} \right|^2}$ denote the distance and the small-scale fading between the user and the $j$-th interfering UAV, $\alpha_I$ denotes the path loss exponent between interfering UAV and the typical user.

Besides, in practical wireless communication systems, obtaining the channel state information (CSI) at the transmitter or receiver is not a trivial problem, which requires the classic pilot-based training process. Therefore, in order to provide more engineering insights, it is assumed that the CSI of UAVs is partly known at the typical user, where only distance information between UAVs and the typical user is required. The signal-to-interference-plus-noise ratio (SINR) of the user-centric strategy will be derived in the following subsection.

\subsection{SINR Analysis}

For the user-centric strategy, since the distance of typical user and its associated UAV is not pre-determined. Therefore, focusing on the typical user, there are two potential cases, namely far user case and near user case, where 1) far user case, i.e., ${r}>r_k$; and 2) near user case, i.e., ${r}<r_k$. We then turn our attention on the SINR analysis of two potential cases.

\emph{(1) Far user case:}

For the far user case, where the serving distance of the typical user is greater than that of the fixed user, the typical user treats the signal from the fixed user as noise, and thus the SINR can be expressed as
\begin{equation}\label{SINR_t far}
SIN{R_{t,far}} = \frac{{{{\left| {{h_{t}}} \right|}^2}r_{t}^{ - \alpha }{P_u}\alpha _{{v}}^2}}{{{\sigma ^2} + {{\left| {{h_{t}}} \right|}^2}r_{t}^{ - \alpha }{P_u}\alpha _w^2 + \sum\limits_{j \in \Psi, d_j>r_t} {{{\left| {{g_{j}}} \right|}^2}{P_u}d_{j}^{ - \alpha_I }} }},
\end{equation}
where ${\sigma ^2}$ denotes the additive white Gaussian noise (AWGN) power, $\alpha _{{v}}^2$ and $\alpha _{{w}}^2$ denote the power allocation factors for the far user and the near user, respectively. Note that $\alpha _{{v}}^2 + \alpha _{{w}}^2=1$ in NOMA communication.

For the far user case, SIC technique is deployed at the fixed user, thereby the fixed user needs to decode the information from the typical user with the following SINR
\begin{equation}\label{SINR_f to t, f near}
SIN{R_{f \to t,far}} = \frac{{{{\left| {{h_{f}}} \right|}^2}R_{k}^{ - \alpha }{P_u}\alpha _{{v}}^2}}{{{\sigma ^2} + {{\left| {{h_{f}}} \right|}^2}R_{k}^{ - \alpha }{P_u}\alpha _w^2 + \sum\limits_{j \in \Psi, d_j>r_t} {{{\left| {{g_{j}}} \right|}^2}{P_u}d_{j}^{ - \alpha_I }} }},
\end{equation}
where $R_k=\sqrt{r_k^2+h^2}$, and $h_{f}$ denotes the channel coefficients for the fixed user.

Once it is decoded successfully, the fixed user will decode its own signal with imperfect SIC coefficient, and the SINR can be expressed as
\begin{equation}\label{SINR_f, f near}
SIN{R_{f,far}} = \frac{{{{\left| {{h_{f}}} \right|}^2}R_{k}^{ - \alpha }{P_u}\alpha _{{w}}^2}}{{{\sigma ^2} + \beta{{\left| {{h_{f}}} \right|}^2}R_{k}^{ - \alpha }{P_u}\alpha _v^2 + \sum\limits_{j \in \Psi, d_j>r_t } {{{\left| {{g_{j}}} \right|}^2}{P_u}d_{j}^{ - \alpha_I }} }},
\end{equation}
where $\beta$ denotes the imperfect SIC coefficient. Since in practice that SIC is not perfect, a fraction $0<\beta<1$ is considered in our model for the user with better channel gain. On the one hand, $\beta=0$ when perfect SIC is assumed, and the near user can perfectly decode the signal intended for the far user. On the other hand, when SIC is failed or there is no corresponding SIC, $\beta=1$.

\emph{(2) Near user case:}

For the near user case, when the typical user has smaller serving distance to the UAV than that of the fixed user, the signal of the typical user can be treated as noise at the fixed user, and thus the SINR of the fixed user can be expressed as
\begin{equation}\label{SINR_f, f far}
SIN{R_{f,near}} = \frac{{{{\left| {{h_{f}}} \right|}^2}R_{k}^{ - \alpha }{P_u}\alpha _{{v}}^2}}{{{\sigma ^2} + {{\left| {{h_{f}}} \right|}^2}R_{k}^{ - \alpha }{P_u}\alpha _w^2 + \sum\limits_{j \in \Psi, d_j> r_t } {{{\left| {{g_{j}}} \right|}^2}{P_u}d_{j}^{ - \alpha_I }} }}.
\end{equation}

The SIC technique can be deployed at the typical user for decoding the signal from the fixed user, and the SINR at the typical user for the near user case can be expressed as
\begin{equation}\label{SINR_t, decode f, t near}
SIN{R_{t \to f,near}} = \frac{{{{\left| {{h_{t}}} \right|}^2}r_{t}^{ - \alpha }{P_u}\alpha _{{v}}^2}}{{{\sigma ^2} + {{\left| {{h_{t}}} \right|}^2}r_{t}^{ - \alpha }{P_u}\alpha _w^2 + \sum\limits_{j \in \Psi, d_j> r_t } {{{\left| {{g_{j}}} \right|}^2}{P_u}d_{j}^{ - \alpha_I }} }}.
\end{equation}

Once the typical user decodes the information from the fixed user successfully, the typical user can decode its own signal with the SINR
\begin{equation}\label{SINR_t, t near}
SIN{R_{t,near}} = \frac{{{{\left| {{h_{t}}} \right|}^2}r_{t}^{ - \alpha }{P_u}\alpha _{{w}}^2}}{{{\sigma ^2} + \beta{{\left| {{h_{t}}} \right|}^2}r_{t}^{ - \alpha }{P_u}\alpha _v^2 + \sum\limits_{j \in \Psi, d_j> r_t } {{{\left| {{g_{j}}} \right|}^2}{P_u}d_{j}^{ - \alpha_I }} }}.
\end{equation}

\subsection{Coverage Probability of the User-centric Strategy}

In the networks considered, we first focus on analyzing the PDF of user distance distributions for paired NOMA users, which will be used for both user-centric strategy and UAV-centric strategy.
\begin{lemma}\label{lemma1:distance distributions}
The UAVs are distributed according to a HPPP with density $\lambda$. It is assumed that the typical user is located at the origin of the disc in the user-centric strategy, or the typical UAV is located at the origin of the disc in the UAV-centric strategy, which is under expectation over HPPP. Thus, the horizontal distance $r$ between the origin and UAVs, follows the distribution
\begin{equation}\label{PDF of serving distance}
{f_r}\left( r \right) = 2\pi \lambda r{e^{ - \pi \lambda {r^2}}},r \ge 0.
\end{equation}
\end{lemma}
Then, we focus on analyzing the user-centric strategy of the proposed framework in order to increase the system fairness. In the user-centric strategy, the user association is based on connecting the nearest UAV to the typical user. As such, the first step is to derive the Laplace transform of interference for the typical user.
\begin{lemma}\label{lemma2:Lapalace transform of interference of typical user}
For the user-centric strategy, and based on the characteristic of stochastic geometry, the interference received at both typical user and fixed user can be recognized as the same. Therefore, the Laplace transform of interference distribution for the paired NOMA users is given by
\begin{equation}\label{laplace transform of typical user in lemma}
\begin{aligned}
\mathcal{L}_t \left( {s} \right) &=
\exp \left( { - \frac{{2\pi {\lambda}}}{\alpha_I }\sum\limits_{i{\rm{ = 1}}}^{m_I} {
   {m_I}  \choose
   i} {{\left( {\frac{{s{P_u}}}{{m_I}}} \right)}^{\delta_I }} {{\left( -1 \right)}^{\delta_I-i }}  B\left( {\frac{-{s{P_u}}}{{m_I}r_{t}^{\alpha_I }};i - \delta_I,1 - {m_I}} \right)} \right),
\end{aligned}
\end{equation}
where $\delta_I  = \frac{2}{\alpha_I }$, $m_I$ denotes the fading parameters between a typical user and interfering UAVs, and $B(;)$ denotes incomplete Beta function.
\begin{proof}
Please refer to Appendix A.
\end{proof}
\end{lemma}

In the case of large-scale networks, the existence of LoS propagations between interfering UAVs at infinity and users is not reasonable. Thus, the minimum received power of inter-cell interference for cellular UAV networks is worth estimating, where the fading parameters between ground users and interfering UAVs equal to one. It is also assumed that the path loss exponent $\alpha_I=4$ because that path loss exponent is normally in the range of 2 to 4, where 2 is for propagation in free space, 4 is for relatively lossy environments and in the case of full specular reflection from the earth surface.

\begin{corollary}\label{coro1:Lapalace transform of interference of typical user in Rayleigh and alpha4}
For the special case that the small scale fading channels between interfering UAVs and users follow Rayleigh fading, thereby $m_I=1$ and $\alpha_I=4$ for the user-centric strategy, the Laplace transform of interference distribution for the both paired NOMA users can be transformed into
\begin{equation}\label{laplace transform of typical user in lemma in Rayleigh}
\begin{aligned}
\mathcal{L}_t \left( {s} \right) & \overset{(a)}{=}
\exp \left( { - \frac{{2\pi {\lambda}{P_u}r_{t}^{2 - \alpha_I }}}{{\alpha_I \left( {1{ - }\delta_I} \right)}}{}_2{F_1}\left( {1,1 - \delta_I;2 - \delta_I; - s{P_u}r_{t}^{ - \alpha_I }} \right)} \right)\\
& \overset{(b)}{=}\exp \left( { - \pi {\lambda }  \sqrt{s{P_u}} {\rm{tan}}^{-1}\left( {\frac{\sqrt{s{P_u}}}{r_{t}^{2 }}} \right)  } \right),
\end{aligned}
\end{equation}
where $(a)$ is resulted from applying $m_I=1$, $(b)$ is obtained by substituting $\alpha_I=4$, and ${}_2{F_1}(;;)$ denotes Gauss hypergeometric function.
\end{corollary}

Then, we focus on the coverage behavior of the user-centric strategy. The fixed power allocation strategy is deployed at the UAV, where the power allocation factors $\alpha_w^2$ and $\alpha_{v}^2$ are constant during transmission. It is assumed that the target rates of the typical user and the fixed user are $R_{t}$ and $R_f$, respectively.
Based on SINR analysis in~\eqref{SINR_t far},~\eqref{SINR_t, decode f, t near} and~\eqref{SINR_t, t near}, the coverage probability of the typical user can be expressed as follows:
\begin{equation}\label{SINR before compare}
\begin{aligned}
&{P_t}(r) = {P_{t,near}}(r)P(r < {r_k}) + {P_{t,far}}(r)P(r > {r_k})\\
& = \Pr \left( {SIN{R_{t \to f,near}} > {\varepsilon _f}, SIN{R_{t,near}} > {\varepsilon _t}} \right) \Pr (r < {r_k}) + \Pr \left( {SIN{R_{t,far}} > {\varepsilon _t}} \right)\Pr (r > {r_k}),
\end{aligned}
\end{equation}
where ${\varepsilon _t} = {2^{{R_t}}} - 1$, ${\varepsilon _f} = {2^{{R_f}}} - 1$, ${P_{t,near}}(r)$ and ${P_{t,far}}(r)$ denote the coverage probability of the typical user for the near user case and the far user case, respectively. $P(r > {r_k})$ and $P(r < {r_k})$ denote the probability of far user case and near user case, respectively.
Therefore, the coverage probability of the typical user for the near user case and far user case is given in following two Lemmas.

\begin{lemma}\label{lemma3:outage of the typical user near user case}
\emph{The coverage probability conditioned on the serving distance of a typical user for the near user case in the user-centric strategy is expressed in closed-form as}
\begin{equation}\label{coverage probability typical user near Lemma}
\begin{aligned}
{P_{t,near}}(r) &= \sum\limits_{n = 0}^{m - 1} {\sum\limits_{p = 0}^n { {
   {n}  \choose
   p}} \frac{{{{( - 1)}^n}}}{{n!}}} {\Lambda _4^n}{\Lambda _5^n}\exp \left( { - m{M_{t*}}{\sigma ^2}r_t^\alpha  - {\Lambda _3} r_t^{2+(\alpha-\alpha_I)(i+a)}} \right) \\
 & \times  r_{t}^{\alpha (1 - j){q_j} +  (2 +(\alpha-\alpha_I)(i+a)- \alpha b){q_b} + \alpha n},
\end{aligned}
\end{equation}
where ${M_t^n} = \frac{{{\varepsilon _t}}}{{{P_u}\left( {\alpha _w^2 - \beta {{\varepsilon _t}}\alpha _v^2} \right)}}$, ${M_{t \to f}} = \frac{{{\varepsilon _f}}}{{{P_u}\left( {\alpha _v^2 -  {{\varepsilon _f}}\alpha _w^2} \right)}}$, ${M_{t*}}=max \left\{ {{{M_t^n},{M_{t \to f}}}} \right\} $, $r_{t}=\sqrt{r^2+h^2}$,\\
 ${\Lambda _3} = \frac{{2\pi m \lambda }}{\alpha_I }\sum\limits_{a = 0}^\infty  {\frac{{{{\left( {{m_I}} \right)}_a}}}{{a!\left( {i - \delta_I  + a} \right)}}} \sum\limits_{i{\rm{ = 1}}}^{{m_I}} {
   {m_I}  \choose
   i} {\left( {\frac{{{M_{t*}}{P_u}}}{{{m_I}}}} \right)^{i + a}}{\left( { - 1} \right)^a}$,
${\Lambda _4^n} = \sum {p!} \prod\limits_{j = 1}^p {\frac{{{{\left( {\left( { - m{M_{t*}}{\sigma ^2}} \right)\prod\limits_{k = 0}^{j - 1} {\left( {1 - k} \right)} } \right)}^{{q_j}}}}}{{{q_j}!{{\left( {j!} \right)}^{{q_j}}}}}}$, and
${\Lambda _5^n} = \sum {(n - p)!} \prod\limits_{b = 1}^{n - p} {\frac{{{{\left( {\left( { - {\Lambda _3}} \right)\prod\limits_{k = 0}^{b - 1} {\left( {\delta_I - k} \right)} } \right)}^{{q_b}}}}}{{{q_b}!{{\left( {b!} \right)}^{{q_b}}}}}}$.
\begin{proof}
Please refer to Appendix B.
\end{proof}
\end{lemma}

For the far user case, note that decoding will succeed if the typical user can decode its own message by treating the signal from the fixed user as noise. The coverage probability conditioned on the serving distance of a typical user for the far user case is calculated in the following Lemma.

\begin{lemma}\label{lemma4:outage of the typical user far user case}
\emph{The coverage probability conditioned on the serving distance of a typical user for the far user case in the user-centric strategy is expressed in closed-form as}
\begin{equation}\label{coverage probability typical user far Lemma}
\begin{aligned}
{P_{t,far}}(r) &= \sum\limits_{n = 0}^{m - 1} {\sum\limits_{p = 0}^n { {
   {n}  \choose
   p}} \frac{{{{( - 1)}^n}}}{{n!}}} {\Lambda _4^f}{\Lambda _5^f}\exp \left( { - m{M_t^f}{\sigma ^2}r_t^\alpha  - {\Lambda _3^f}r_t^{2 +(\alpha-\alpha_I)(i+a) }} \right)\\
   & \times r_{t}^{\alpha (1 - j){q_j} +  (2 +(\alpha-\alpha_I)(i+a) - \alpha b){q_b} + \alpha n},
   \end{aligned}
\end{equation}
where ${M_{t}^f} = \frac{{{\varepsilon _t}}}{{{P_u}\left( {\alpha _v^2 -  {{\varepsilon _t}}\alpha _w^2} \right)}}$,
${\Lambda _3^f} = \frac{{2\pi m \lambda }}{\alpha_I }\sum\limits_{a = 0}^\infty  {\frac{{{{\left( {{m_I}} \right)}_a}}}{{a!\left( {i - \delta_I  + a} \right)}}} \sum\limits_{i{\rm{ = 1}}}^{{m_I}} { {
   {m_I}  \choose
   i} } {\left( {\frac{{{M_t^f}{P_u}}}{{{m_I}}}} \right)^{i + a}}{\left( { - 1} \right)^a}$,\\
${\Lambda _4^f} = \sum {p!} \prod\limits_{j = 1}^p {\frac{{{{\left( {\left( { - m{M_t^f}}{\sigma ^2} \right)\prod\limits_{k = 0}^{j - 1} {\left( {1 - k} \right)} } \right)}^{{q_j}}}}}{{{q_j}!{{\left( {j!} \right)}^{{q_j}}}}}}$, and
${\Lambda _5^f} = \sum {(n - p)!} \prod\limits_{b = 1}^{n - p} {\frac{{{{\left( {\left( { - {\Lambda _3^f}} \right)\prod\limits_{k = 0}^{b - 1} {\left( {\delta - k} \right)} } \right)}^{{q_b}}}}}{{{q_b}!{{\left( {b!} \right)}^{{q_b}}}}}}$.
\begin{proof}
Based on the SINR analysis in \eqref{SINR_t far}, and following the similar procedure in Appendix~B, with interchanging ${M_{t*}}$ with ${M_{t}^f}$, we can obtain the desired result in \eqref{coverage probability typical user far Lemma}. Thus, the proof is complete.
\end{proof}
\end{lemma}

\begin{remark}\label{user-centric}
The derived results in \eqref{coverage probability typical user near Lemma} and \eqref{coverage probability typical user far Lemma} demonstrate that the coverage probability of a typical user is determined by imperfect SIC coefficient, the target rate of itself, fading parameter $m$ of the small scale fading channels and the distance of the fixed user served by the same UAV.
\end{remark}

\begin{remark}\label{user-centric power allocation}
Inappropriate power allocation such as, $\alpha _v^2 -  {{\varepsilon _t}}\alpha _w^2<0$ and ${\alpha _w^2 - \beta {{\varepsilon _t}}\alpha _v^2}<0$, will lead to the coverage probability always being zero.
\end{remark}

Based on {\bf{Lemma \ref{lemma3:outage of the typical user near user case}}} and {\bf{Lemma \ref{lemma4:outage of the typical user far user case}}}, the coverage probability of the typical user in the user-centric strategy can be calculated in the following Theorem.

\begin{theorem}\label{theorem 1 coverage probability of typical user UserCentric}
\emph{The exact expression of the coverage probability for the typical user is expressed as}
\begin{equation}\label{coverage of typical user expression}
\begin{aligned}
{P_{t}} = \int\limits_0^{r_k} {P_{t,near}}(r)  {f_r}\left( r \right)dr+ \int\limits_{r_k}^{\infty} {P_{t,far}}(r)  {f_r}\left( r \right)dr,
\end{aligned}
\end{equation}
where ${P_{t,near}}(r)$ is given in \eqref{coverage probability typical user near Lemma}, ${P_{t,far}}(r)$ is given in \eqref{coverage probability typical user far Lemma}, and ${f}\left( r \right)$ is given in \eqref{PDF of serving distance}.
\end{theorem}

\begin{remark}\label{remark 3: typical user coverage}
Based on the result in~\eqref{coverage of typical user expression}, the coverage probability of the typical user is dependent on the distance of the fixed user in the user-centric strategy.
\end{remark}

In order to provide more insights for UAV assisted cellular networks, the coverage probability of the typical user is also derived in the OMA assisted UAV cellular networks, i.e., TDMA. The typical user and fixed user follow the same distance distributions and small-scale fading channels in the OMA assisted cellular UAV networks. The OMA benchmark adopted in this article is that by dividing the two users in equal time/frequency slots.

\begin{corollary}\label{corollary2: typical user OMA case}
\emph{The coverage probability conditioned on the serving distance of a typical user for the OMA assisted UAV cellular networks in the user-centric strategy is expressed in closed-form as}
\begin{equation}\label{coverage probability typical user far Lemma OMA case}
\begin{aligned}
 {P_{cov,t,o}}(r) &= \sum\limits_{n = 0}^{m - 1} {\sum\limits_{p = 0}^n { {
   {n}  \choose
   p}} \frac{{{{( - 1)}^n}}}{{n!}}} {\Lambda _4^o}{\Lambda _5^o}\exp \left( { - m{M_t^o}{\sigma ^2}r_t^\alpha  - {\Lambda _3^o}r_t^{2+(\alpha-\alpha_I)(i+a)}} \right) \\
   &\times r_{t}^{\alpha (1 - j){q_j} +  (2 +(\alpha-\alpha_I)(i+a) -  \alpha b){q_b} + \alpha n},
\end{aligned}
\end{equation}
where ${M_t^o} = \frac{{{\varepsilon _t^o}}}{{{P_u}}}$, ${\varepsilon _t^o} = {2^{{2R_t}}} - 1$, ${\Lambda _3^o} = \frac{{2\pi m \lambda }}{\alpha_I }\sum\limits_{a = 0}^\infty  {\frac{{{{\left( {{m_I}} \right)}_a}}}{{a!\left( {i - \delta_I  + a} \right)}}} \sum\limits_{i{\rm{ = 1}}}^{{m_I}} { {
   {m_I}  \choose
   i} } {\left( {\frac{{{M_t^o}{P_u}}}{{{m_I}}}} \right)^{i + a}}{\left( { - 1} \right)^a}$,\\
${\Lambda _4^o} = \sum {p!} \prod\limits_{j = 1}^p {\frac{{{{\left( {\left( { - m{M_t^o}}{\sigma ^2} \right)\prod\limits_{k = 0}^{j - 1} {\left( {1 - k} \right)} } \right)}^{{q_j}}}}}{{{q_j}!{{\left( {j!} \right)}^{{q_j}}}}}}$, and
${\Lambda _5^o} = \sum {(n - p)!} \prod\limits_{b = 1}^{n - p} {\frac{{{{\left( {\left( { - {\Lambda _3}} \right)\prod\limits_{k = 0}^{b - 1} {\left( {\delta_I - k} \right)} } \right)}^{{q_b}}}}}{{{q_b}!{{\left( {b!} \right)}^{{q_b}}}}}}$.
\begin{proof}
Following the similar procedure in Appendix B, with interchanging ${M_{t}^f}$ with ${M_{t}^o}$, we can obtain the desired result in \eqref{coverage probability typical user far Lemma OMA case}. Thus, the proof is complete.
\end{proof}
\end{corollary}

\section{UAV-centric Strategy for Offloading Actions}

\begin{figure*}[t!]
\centering
\includegraphics[width =3in]{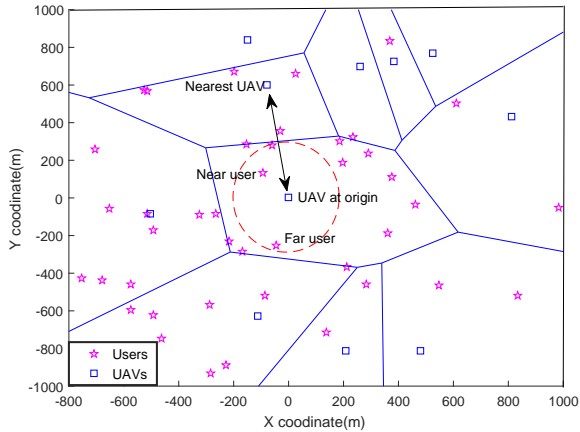}
\caption{Top view of the UAV-centric strategy cellular networks.}
\label{Example of UAV-Centric}
\vspace{-0.3in}
\end{figure*}

In conventional BS communications, the BSs are distributed in order to cover all the ground, whereas UAV communications mainly focus on providing access services to support BSs in the hotspot areas of the dense networks, i.e., airports or resorts, where most users are located in the lounge~\cite{Lyu_UAV_hotspots}. Based on the insights of~\cite{NOMA_downlink_cellular}, where the serving area can be considered as a regular disc, another strategy considered in this article is the UAV-centric strategy, where paired NOMA users are located inside the coverage disc as shown in Fig.~\ref{Example of UAV-Centric}. It is also worth noting that the locations of UAVs are properly selected to serve terrestrial users in the hotspot areas based on user density in the UAV-centric strategy. Based on the insights of poisson cluster process (PCP), the users are located in multiple small clusters in practice.

For the UAV-centric strategy, a UAV is located at the original point, which becomes the typical UAV serving users in the typical cell. Therefore, it is assumed that the distance between the UAV at the origin and the nearest UAV is $R$, and the potential paired NOMA users are located in the coverage area within the radius $R/2$. In the UAV-centric strategy, user pairing strategy is determined by the connected UAV, where all the users in the coverage disc are connected to the UAV.
In the user association, for simplicity, we assume that there are two users, near user $w$ and far user $v$, have accessed to the UAV at the origin to perform NOMA. It is assumed that the users are uniformly located, which is according to HPPP, denoted by $\Psi_u$ and it is associated with the density $\lambda_u$, within large ring and small disc with radius $R/2$ and $R/4$, respectively. By doing so, NOMA technique can be performed without accurate CSI. 

\subsection{SINR Analysis}

For the UAV-centric strategy, the distances between the interfering UAVs and the users are more complicated. For notational simplicity, the location of the $j$-th interfering UAV is denoted by $y_j$, where ${y_j} \in \Psi$. The locations of the users are conditioned on the locations of their cluster heads (UAVs). As such,
the SINR of the far user $v$ can be derived as
\begin{equation}\label{SINR_v}
SIN{R_{v}} = \frac{{{{\left| {{h_{v}}} \right|}^2}d_{v}^{ - \alpha }{P_u}\alpha _{{v}}^2}}{{{\sigma ^2} + {{\left| {{h_{v}}} \right|}^2}d_{v}^{ - \alpha }{P_u}\alpha _w^2 + \sum\limits_{j \in \Psi } {{{\left| {{g_{j}}} \right|}^2}{P_u}d_{j}^{ - \alpha_I }} }},
\end{equation}
where ${{\left| {{h_{v}}} \right|}^2}$ and $d_{v}$ denote the small scale fading coefficient and the distance between the far user and the UAV, ${{\left| {{g_{j}}} \right|}^2}$ and $d_{j}$ denote the small scale fading coefficient and the distance between $j$-th interfering UAV and the user, respectively.

The near user $w$ will first decode the signal of the far user $v$ with the following SINR
\begin{equation}\label{SINR_w tov}
SIN{R_{w \to v}} = \frac{{{{\left| {{h_w}} \right|}^2}d_w^{ - \alpha }{P_u}\alpha _{v}^2}}{{{\sigma ^2} + \beta {{\left| {{h_w}} \right|}^2}d_w^{ - \alpha }{P_u}\alpha _{w}^2 + \sum\limits_{j \in \Psi } {{{\left| {{g_{j}}} \right|}^2}{P_u}d_{j}^{ - \alpha_I }} }},
\end{equation}
where ${{\left| {{h_{w}}} \right|}^2}$ and $d_{w}$ denote the small scale fading coefficient and the distance between the near user and the UAV.
If the signal of the $v$-th user can be decoded successfully, the $w$-th user then decodes its own signal. As such, the SINR at the $w$-th user can be expressed as
\begin{equation}\label{SINR_w}
SIN{R_w} = \frac{{{{\left| {{h_w}} \right|}^2}d_w^{ - \alpha }{P_u}\alpha _{w}^2}}{{{\sigma ^2} + \beta {{\left| {{h_w}} \right|}^2}d_w^{ - \alpha }{P_u}\alpha _{v}^2 + \sum\limits_{j \in \Psi } {{{\left| {{g_{j}}} \right|}^2}{P_u}d_{j}^{ - \alpha_I }} }}.
\end{equation}

\subsection{Coverage Probability of the UAV-centric Strategy}
Consider a disk centered at the origin with the radius $R/2$, which has shown in Fig.~\ref{Example of UAV-Centric}. In order to deploy NOMA protocol, we separate the disc to two parts equally, the small disc with radius $R/4$ and the ring with radius from $R/4$ to $R/2$, to serve paired NOMA users. It is assumed that the near users and the far users are located in the small disc and ring, respectively. Focusing on the typical cell, where a UAV is located at the origin, the PDF of distance for the near users conditioned on serving distance $R$, follows
\begin{equation}\label{condition pdf near user}
{f_w} \left( {r\left| R \right.} \right) = \frac{{32r}}{{{R^2}}},0 \le r \le l_1,
\end{equation}
where ${l_1} = { {\frac{R}{4}} } $.

The PDF of far users can be obtained by
\begin{equation}\label{condition pdf far user}
{f_v} \left( {r\left| R \right.} \right) = \frac{{32r}}{{3{R^2}}}, l_1 \le r \le  l_2,
\end{equation}
where ${l_2} = { {\frac{R}{2}}} $.

In order to derive the system performance, the Laplace transform of UAV interferences needs to be derived. We calculate the Laplace transform of inter-cell interference for the paired users in the following Lemma.
\begin{lemma}\label{Lapalace transform of interference of near user}
For the UAV-centric strategy, the Laplace transform of interference distribution conditioned on the serving distance $R$ for paired user is given by
\begin{equation}\label{laplace transform of UAV in lemma}
\begin{aligned}
\mathcal{L}_U \left( {s\left| R \right.} \right) &= \exp \left({ - \frac{{{l_I}}}{R}\left( {1 - {{\left( {1 + \frac{{S{P_u}}}{{{m_I}l_I^{\alpha_I} }}} \right)}^{ - {m_I}}}} \right)    }\right)  \\
& \times \exp \left( { - \frac{{2\pi {\lambda }}}{\alpha_I }\sum\limits_{i{\rm{ = 1}}}^{m_I} {{
   {m_I}  \choose
   i}} {{\left( {\frac{{s{P_u}}}{{m_I}}} \right)}^{\delta_I}}(-1)^{(\delta_I-i)}B\left( {\frac{{-s{P_u}l_I^{ - \alpha_I }}}{{m_I}};i - \delta_I,1-{m_I}} \right)} \right).
\end{aligned}
\end{equation}
where ${l_I} = \sqrt {{R^2} + {h^2}}$.
\begin{proof}
Please refer to Appendix C.
\end{proof}
\end{lemma}

It is also worth noting that for the NLoS case, the small-scale fading between users and interfering UAVs can be considered as Rayleigh fading. Thus, the Laplace transform can be further obtained in the following Corollary.
\begin{corollary}\label{Lapalace transform Rayleigh of near user}
For the NLoS scenario, the Laplace transform of interference distribution conditioned on the serving distance $R$ is given by
\begin{equation}\label{laplace transform of near user in lemma of Rayleigh}
\mathcal{L}_U \left( {s\left| R \right.} \right)  = \exp \left( { - \frac{{{l_I}}}{R}\left( {\frac{{S{P_u}}}{{l_I^{\alpha_I}  + S{P_u}}}} \right)  } \right) \exp \left( { - \frac{{2\pi {\lambda }{P_u}l_I^{2 - \alpha_I }}}{{\alpha_I \left( {1{\rm{ - }}\delta_I } \right)}}{}_2{F_1}\left( {1,1 - \delta_I ;2 - \delta_I ; - s{P_u}l_I^{ - \alpha_I }} \right)} \right).
\end{equation}
\end{corollary}

Then, we focus on the coverage behavior of paired NOMA users in the UAV-centric strategy. In the UAV-centric strategy, the coverage probability is more complicated than the user-centric strategy due to the fact that the interfering UAV located at distance $R$ is necessary to evaluate separately.
It is assumed that the target rates of user $w$ and user $v$ are $R_{w}$ and $R_{v}$, respectively. Therefore, the coverage probability of the $w$-th user is given in the following Lemma.
\begin{lemma}\label{lemma5:outage of the near user conditioned on radius}
\emph{ The closed-form expression of the coverage probability conditioned on serving distance for the near user is expressed as}
\begin{equation}\label{coverage probability conditioned on radius Lemma near user UAV-centric}
\begin{aligned}
&{P_{cov,w}}\left( {r\left| R \right.} \right) = \sum\limits_{n = 0}^{m - 1} {\sum\limits_{k = 0}^n {\sum\limits_{l = 0}^k {\frac{{{{( - 1)}^n}r_w^{\alpha n}}}{{l!(k - l)!(n - k)!}}} } } {\Theta _3}{\Theta _4}{\Theta _5}\\
&\times \exp \left( { - m{M_{w*}}{\sigma ^2}r_w^\alpha  - {\Theta _1}r_w^{\alpha (i + a)} - \frac{{m{l_I}}}{R} + {\Theta _2}r_w^{\alpha U}} \right) r_w^{\alpha (1 - j){q_j} + \alpha (i + a - g){q_g} + \alpha n + \alpha (U - b){q_u}},
\end{aligned}
\end{equation}
where ${M_w} = \frac{{{\varepsilon _w}}}{{{P_u}\left( {\alpha _w^2 - \beta {\varepsilon _w}\alpha _v^2} \right)}}$, ${M_v} = \frac{{{\varepsilon _v}}}{{{P_u}\left( {\alpha _v^2 - {\varepsilon _v}\alpha _w^2} \right)}}$, ${\varepsilon _w} = {2^{{R_w}}} - 1$, ${\varepsilon _v} = {2^{{R_v}}} - 1$, ${M_{w*}} = \max \left\{ {{M_w},{M_v}} \right\}$, $r_w=\sqrt{r^2+h^2}$,
 ${\Theta _1} = \pi m \delta_I {\lambda}\sum\limits_{i{\rm{ = 1}}}^{{m_I}} { {
   {m_I}  \choose
   i} } {\left( { - 1} \right)^{\delta_I - 1}}\sum\limits_{a = 0}^\infty  {\frac{{{{\left( {{m_I}} \right)}_a}}}{{a!\left( {i - \delta_I + a} \right)}}} {\left( {\frac{{{M_{w*}}{P_u}}}{{{m_I}}}} \right)^{i + a}}l_I^{{\rm{ - }}\alpha_I \left( {i - \delta_I  + a} \right)}$, \\
${\Theta _2} = \frac{{m{l_I}}}{R}\sum\limits_{U = 0}^\infty  {{{( - 1)}^U}C_{{m_I} + U + 1}^U} {\left( {\frac{{{M_{w*}}{P_u}}}{{l_I^{\alpha_I} {m_I}}}} \right)^U}$,
${\Theta _3} = \sum {(n - k)!} \prod\limits_{j = 1}^{n - k} {\frac{{{{\left( {\left( { - m{M_{w*}}{\sigma ^2}} \right)\prod\limits_{p = 0}^{j - 1} {\left( {1 - p} \right)} } \right)}^{{q_j}}}}}{{{q_j}!{{\left( {j!} \right)}^{{q_j}}}}}}$, \\
${\Theta _4} = \sum {(k - l)!} \prod\limits_{b = 1}^{k - l} {\frac{{{{\left( {\left( { - {\Theta _2}} \right)\prod\limits_{p = 0}^{b - 1} {\left( {U - p} \right)} } \right)}^{{q_u}}}}}{{{q_u}!{{\left( {j!} \right)}^{{q_u}}}}}} $, and ${\Theta _5} = \sum {l!} \prod\limits_{g = 1}^{l} {\frac{{{{\left( {\left( { - {\Theta _1}} \right)\prod\limits_{p = 0}^{g - 1} {\left( {i+a-g} \right)} } \right)}^{{q_g}}}}}{{{q_g}!{{\left( {j!} \right)}^{{q_g}}}}}} $.
\begin{proof}
Please refer to Appendix D.
\end{proof}
\end{lemma}

Similar to \textbf{Lemma~\ref{lemma5:outage of the near user conditioned on radius}}, the coverage probability of the far user can be derived in the following Lemma.
\begin{lemma}\label{lemma7:outage of the far user conditioned on radius}
\emph{ The closed-form expression of the coverage probability conditioned on serving distance for the far user is expressed as}
\begin{equation}\label{coverage probability conditioned on radius Lemma_far_UAV-centric}
\begin{aligned}
&{P_{cov,v}}\left( {r\left| R \right.} \right) = \sum\limits_{n = 0}^{m - 1} {\sum\limits_{k = 0}^n {\sum\limits_{l = 0}^k {\frac{{{{( - 1)}^n}r_v^{\alpha n}}}{{l!(k - l)!(n - k)!}}} } } {\Theta _{3,v}}{\Theta _{4,v}}{\Theta _{5,v}}\\
&\times \exp \left( { - m{M_{v}}{\sigma ^2}r_v^\alpha  - {\Theta _{1,v}}r_v^{\alpha (i + a)} - \frac{{m{l_I}}}{R} + {\Theta _{2,v}}r_v^{\alpha U}} \right) r_v^{\alpha (1 - j){q_j} + \alpha (i + a - g){q_g} + \alpha n + \alpha (U - b){q_u}},
\end{aligned}
\end{equation}
where $r_v=\sqrt{r^2+h^2}$, ${\Theta _{1,v}} = \pi m \delta_I {\lambda}\sum\limits_{i{\rm{ = 1}}}^{{m_I}} { {
   {m_I}  \choose
   i} } {\left( { - 1} \right)^{\delta_I - 1}}\sum\limits_{a = 0}^\infty  {\frac{{{{\left( {{m_I}} \right)}_a}}}{{a!\left( {i - \delta_I + a} \right)}}} {\left( {\frac{{{M_{v}}{P_u}}}{{{m_I}}}} \right)^{i + a}}l_I^{{\rm{ - }}\alpha_I \left( {i - \delta_I  + a} \right)}$, \\
${\Theta _{2,v}} = \frac{{m{l_I}}}{R}\sum\limits_{U = 0}^\infty  {{{( - 1)}^U}C_{{m_I} + U + 1}^U} {\left( {\frac{{{M_{v}}{P_u}}}{{l_I^{\alpha_I} {m_I}}}} \right)^U}$,
${\Theta _{3,v}} = \sum {(n - k)!} \prod\limits_{j = 1}^{n - k} {\frac{{{{\left( {\left( { - m{M_{v}}{\sigma ^2}} \right)\prod\limits_{p = 0}^{j - 1} {\left( {1 - p} \right)} } \right)}^{{q_j}}}}}{{{q_j}!{{\left( {j!} \right)}^{{q_j}}}}}}$, \\
${\Theta _{4,v}} = \sum {(k - l)!} \prod\limits_{b = 1}^{k - l} {\frac{{{{\left( {\left( { - {\Theta _{2,v}}} \right)\prod\limits_{p = 0}^{b - 1} {\left( {U - p} \right)} } \right)}^{{q_u}}}}}{{{q_u}!{{\left( {j!} \right)}^{{q_u}}}}}} $, and ${\Theta _{5,v}} = \sum {l!} \prod\limits_{g = 1}^{l} {\frac{{{{\left( {\left( { - {\Theta _{1,v}}} \right)\prod\limits_{p = 0}^{g - 1} {\left( {i+a-g} \right)} } \right)}^{{q_g}}}}}{{{q_g}!{{\left( {j!} \right)}^{{q_g}}}}}} $.
\begin{proof}
Similar to Appendix D, the derivation in \eqref{coverage probability conditioned on radius Lemma_far_UAV-centric} can be readily proved.
\end{proof}
\end{lemma}

Then, the coverage probability of paired NOMA users in the UAV-centric strategy can be derived in the following Theorem.
\begin{theorem}\label{theorem 1 coverage probability}
\emph{Based on {\bf{Lemma \ref{lemma5:outage of the near user conditioned on radius}}} and {\bf{Lemma \ref{lemma7:outage of the far user conditioned on radius}}}, the exact expressions of the coverage probability for the paired NOMA users can be expressed as}
\begin{equation}\label{coverage of near user expression}
\begin{aligned}
{P_{cov,w}} = \int\limits_0^\infty  {\int\limits_0^{{l_1}} {{P_{cov,w}}} \left( {\left. r \right|R} \right){f_w}\left( {r\left| R \right.} \right)dr} {f_r}\left( R \right)dR,
\end{aligned}
\end{equation}
and
\begin{equation}\label{coverage of far user expression}
\begin{aligned}
{P_{cov,v}} = \int\limits_0^\infty  {\int\limits_{{l_1}}^{{l_2}} {{P_{cov,v}}} \left( {\left. r \right|R} \right){f_v}\left( {r\left| R \right.} \right)dr} {f_r}\left( R \right)dR,
\end{aligned}
\end{equation}
where $l_1=\frac{R}{4}$, $l_2=\frac{R}{2}$, ${P_{cov,w}} \left( {\left. r \right|R} \right)$ is given in \eqref{coverage probability conditioned on radius Lemma near user UAV-centric}, ${P_{cov,w}} \left( {\left. r \right|R} \right)$ is given in \eqref{coverage probability conditioned on radius Lemma_far_UAV-centric}, ${{f_w}\left( {r\left| R \right.} \right)}$ is given in \eqref{condition pdf near user}, ${{f_v}\left( {r\left| R \right.} \right)}$ is given in~\eqref{condition pdf far user}, and ${f_r}\left( R \right)$ is given in \eqref{PDF of serving distance}.
\begin{proof}
By utilizing the PDF in~\eqref{condition pdf near user}, the coverage probability of the near user conditioned on the serving distance can be obtained by
\begin{equation}\label{outage second expression appendix B}
\begin{aligned}
{P_{cov,w}(R)} =  {\int\limits_0^{{l_1}} {{P_{cov,w}}} \left( {\left. r \right|R} \right){f_w}\left( {r\left| R \right.} \right)dr},
\end{aligned}
\end{equation}

The overall coverage probability can be derived by the serving distance of UAV assisted cellular networks, which can be expressed as
\begin{equation}\label{overall coverage probability expression}
{P_{cov,w}} = \int\limits_0^\infty  {{P_{cov,w}(R)}} {f_r}\left( R \right)dR,
\end{equation}

Plugging \eqref{PDF of serving distance} into \eqref{overall coverage probability expression}, and after some mathematical manipulations, the coverage probability of the near user can be obtained. Thus, the proof is complete.
\end{proof}
\end{theorem}

In order to provide more engineering insights, the coverage probability for the near user in the OMA assisted UAV-centric strategy is also derived in the following Corollary. Similar to \textbf{Corollary~\ref{corollary2: typical user OMA case}}, we also use TDMA to illustrate the coverage performance for OMA assisted UAV-centric strategy.

\begin{corollary}\label{corollary4: near user OMA case}
\emph{The coverage probability conditioned on the serving distance of the near user for the OMA enhanced UAV-centric strategy is expressed in closed-form as}
\begin{equation}\label{UAVCentric OMA}
\begin{aligned}
&{P_{cov,w}^o}\left( {r\left| R \right.} \right) = \sum\limits_{n = 0}^{m - 1} {\sum\limits_{k = 0}^n {\sum\limits_{l = 0}^k {\frac{{{{( - 1)}^n}r_w^{\alpha n}}}{{l!(k - l)!(n - k)!}}} } } {\Theta _3^o}{\Theta _4^o}{\Theta _5^o}\\
&\times \exp \left( { - m{M_{w}^o}{\sigma ^2}r_w^\alpha  - {\Theta _1}r_w^{\alpha (i + a)} - \frac{{m{l_I}}}{R} + {\Theta _2}r_w^{\alpha U}} \right) r_w^{\alpha (1 - j){q_j} + \alpha (i + a - g){q_g} + \alpha n + \alpha (U - b){q_u}},
\end{aligned}
\end{equation}
where ${M_w^o} = \frac{{{\varepsilon _w^o}}}{{{P_u}}}$, ${\varepsilon _w^o} = {2^{{2R_w}}} - 1$, ${\Theta _1^o} = \pi m \delta_I {\lambda}\sum\limits_{i{\rm{ = 1}}}^{{m_I}} { {
   {m_I}  \choose
   i} } {\left( { - 1} \right)^{\delta_I - 1}}\sum\limits_{a = 0}^\infty  {\frac{{{{\left( {{m_I}} \right)}_a}}}{{a!\left( {i - \delta_I + a} \right)}}} {\left( {\frac{{{M_{w}^o}{P_u}}}{{{m_I}}}} \right)^{i + a}}l_I^{{\rm{ - }}\alpha_I \left( {i - \delta_I  + a} \right)}$, \\
${\Theta _2^o} = \frac{{m{l_I}}}{R}\sum\limits_{U = 0}^\infty  {{{( - 1)}^U}C_{{m_I} + U + 1}^U} {\left( {\frac{{{M_{w}^o}{P_u}}}{{l_I^{\alpha_I} {m_I}}}} \right)^U}$,
${\Theta _3^o} = \sum {(n - k)!} \prod\limits_{j = 1}^{n - k} {\frac{{{{\left( {\left( { - m{M_{w}^o}{\sigma ^2}} \right)\prod\limits_{p = 0}^{j - 1} {\left( {1 - p} \right)} } \right)}^{{q_j}}}}}{{{q_j}!{{\left( {j!} \right)}^{{q_j}}}}}}$, \\
${\Theta _4^o} = \sum {(k - l)!} \prod\limits_{b = 1}^{k - l} {\frac{{{{\left( {\left( { - {\Theta _2^o}} \right)\prod\limits_{p = 0}^{b - 1} {\left( {U - p} \right)} } \right)}^{{q_u}}}}}{{{q_u}!{{\left( {j!} \right)}^{{q_u}}}}}} $, and ${\Theta _5^o} = \sum {l!} \prod\limits_{g = 1}^{l} {\frac{{{{\left( {\left( { - {\Theta _1^o}} \right)\prod\limits_{p = 0}^{g - 1} {\left( {i+a-g} \right)} } \right)}^{{q_g}}}}}{{{q_g}!{{\left( {j!} \right)}^{{q_g}}}}}} $.
\begin{proof}
Following the similar procedure in Appendix D, with interchanging ${M_{w}^f}$ with ${M_{w}^o}$, we can obtain the desired result in~\eqref{UAVCentric OMA}. Thus, the proof is complete.
\end{proof}
\end{corollary}

\section{Numerical Studies}

In this section, numerical results are provided to facilitate the performance evaluation of NOMA assisted UAV cellular networks. Monte Carlo simulations are conducted to verify analytical results. In the considered network, it is assumed that the power allocation factors are $\alpha_{v}^2=0.6$ for the far user and $\alpha_{w}^2=0.4$ for the near user. The path loss exponent of interference links $\alpha_I$ is set to 4, and the path loss exponent of the desired transmission is smaller than 4. The height of the UAV is fixed to 100 meters. In Monte Carlo simulations, it is not possible to simulate a real infinite distribution for UAVs. Hence, the UAVs are distributed in a disc, and the radius of the disc is $10000$m. The bandwidth of the downlink transmission is set as $BW=300$ kHz, and the power of AWGN noise is set as $\sigma^2= −-174+ 10 {\rm{log}}_{10}(BW)$ dBm. The UAV density $\lambda=\frac{1}{500^2\pi}$. It is also worth noting that LoS and NLoS scenarios are indicated by the Nakagami fading parameter $m$, where $m = 1$ for NLoS scenarios (Rayleigh fading) and $m >1$ for LoS scenarios. Without loss of generality, we use $m=2$ to represent LoS scenario in Section IV.

\subsection{User-centric strategy}

\begin{figure*}[t!]
\centering
\subfigure[Coverage probability of user-centric NOMA versus transmit power in NLoS scenario with different path loss exponent, where the fading parameters $m=1$ and $m_I=1$.]{\label{Outage_UserCentric with m=1}
\includegraphics[width =2.8in]{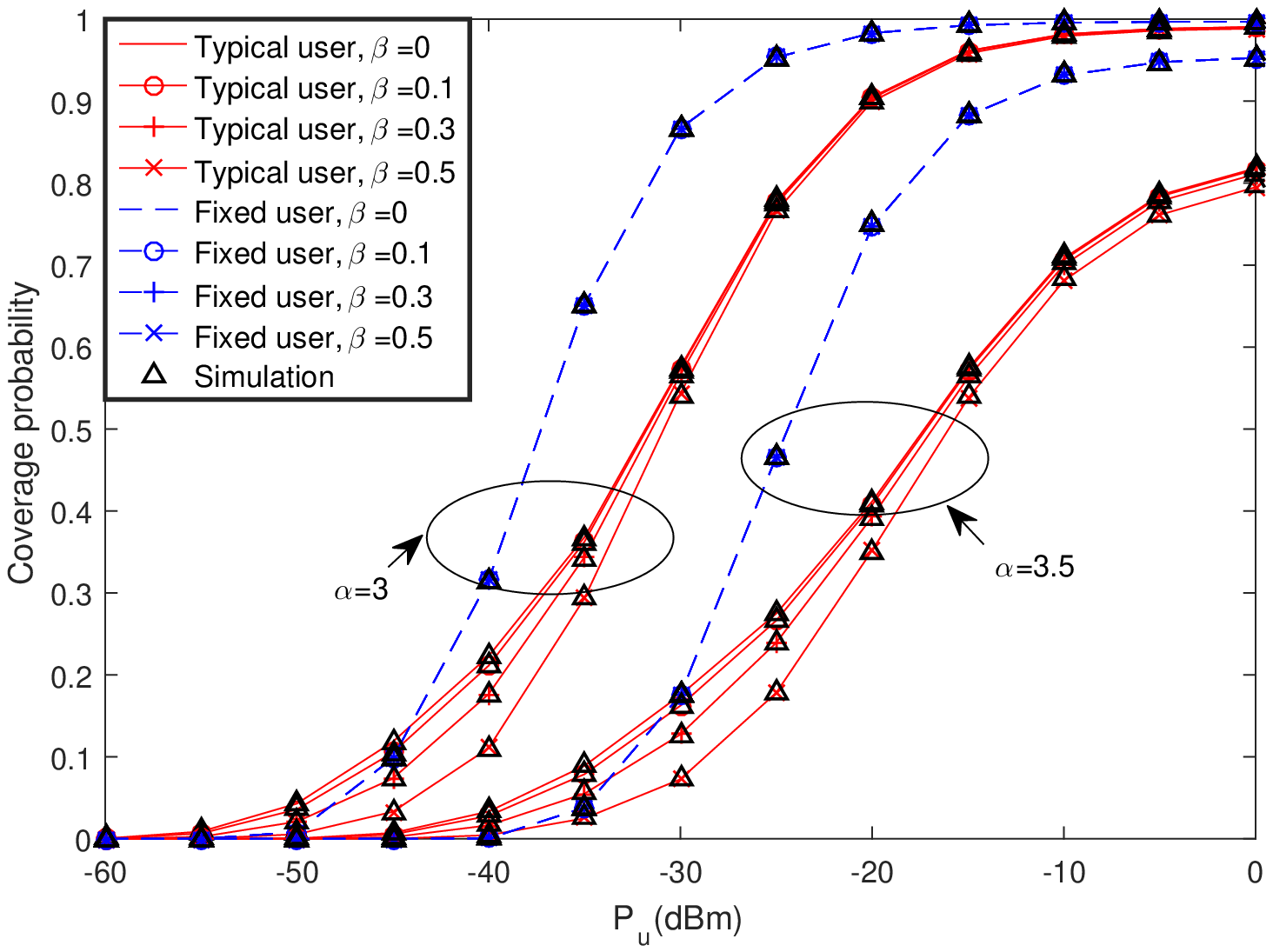}}
\subfigure[Coverage probability of user-centric NOMA versus transmit power in both NLoS and LoS scenarios, where the fading parameters $m=2$ and $m_I=1$. The path loss exponent of desire link is set to be $\alpha=3$.]{\label{Outage_UserCentric with m=2}
\includegraphics[width= 2.8in]{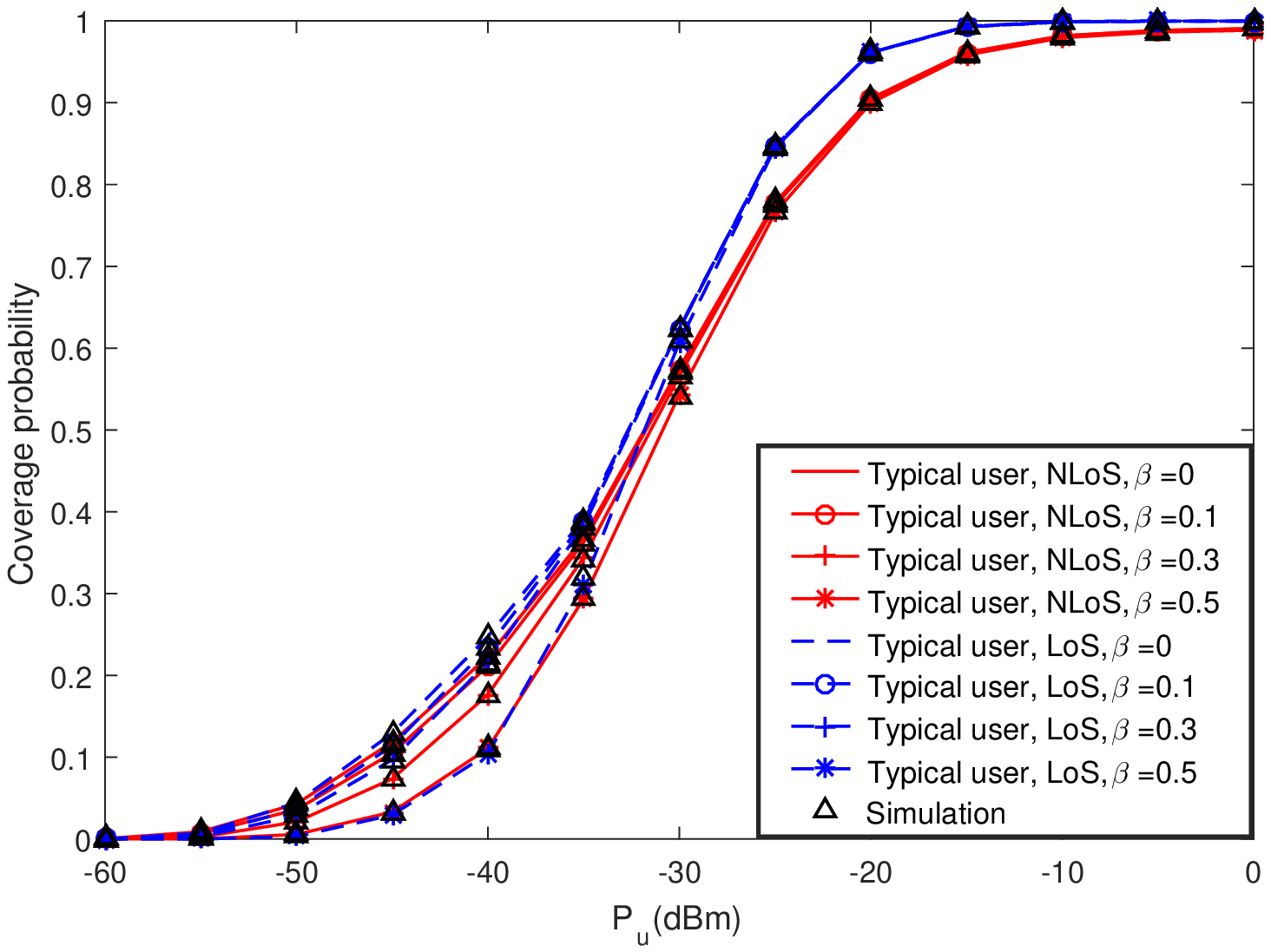}}
\caption{Coverage probability of paired NOMA users versus the power of UAV in the user-centric strategy, with target rate $R_{t}=1$ BPCU and $R_f=0.5$ BPCU. The horizontal distance of the fixed user is 300m. The exact results of NOMA are calculated from \eqref{coverage of typical user expression}.}
\label{Fig1:Outage_UserCetric}
\vspace{-0.3in}
\end{figure*}

First, we evaluate the coverage performance of downlink NOMA users in the user-centric strategy. In Fig.~\ref{Outage_UserCentric with m=1}, for a given set of the distance of fixed users, the solid curves and dashed curves are the coverage probability for typical users and fixed users, respectively. We can see that, as the power of UAV increases, the coverage ceilings, which are the maximum coverage probability for the proposed networks, of both typical users and fixed NOMA users occur. This is due to the fact that, as the higher power level of interfering UAVs is deployed, the received SINR decreases dramatically. It is observed that as imperfect SIC coefficient $\beta$ increases, the coverage probability of typical users decreases, which indicates that the performance of NOMA assisted UAV communication can be effectively improved by decreasing the imperfect SIC coefficient. For example, in the case of $\beta=\frac{2}{3}$, the power residual from imperfect SIC is greater than the power of near users, i.e., $\alpha_w^2<\alpha_v^2\beta$. We can also see that in the case of $\beta=0, 0.1, 0.3, 0.5$, the coverage probabilities of fixed users are the same. This is due to the fact that the imperfect SIC is the critical component of typical users, whereas the imperfect SIC has no effect for fixed users in the case $R_f=0.5$ bits per channel use (BPCU). As we can see in the figure, the outage of typical users occurs more frequently than fixed users. This is due to the fact that the choice of power allocation factors and the distance of fixed users. Note that the simulation results and analytical results match perfectly in Fig.~\ref{Outage_UserCentric with m=1}, which demonstrate the accuracy of the developed analytical results.

Fig. \ref{Outage_UserCentric with m=2} shows the coverage probability achieved by typical users in both NLoS and LoS scenario. In order to better illustrate the performance affected by the LoS transmission, the NLoS case is also shown in the figure as a benchmark for comparison. In Fig.~\ref{Outage_UserCentric with m=2}, we can see that higher fading parameter $m$ would result in reduced outage probability for different UAV power levels and different imperfect SIC coefficients. This is because that the LoS link between the UAV and users provides higher received power
level. It is also worth noting that for the UAV cellular networks, the proposed network is not in need of a larger UAV power for increasing the coverage probability due to the fact that the coverage ceiling occurs in the high SNR regime.

\begin{figure*}[t!]
\centering
\includegraphics[width =2.8in]{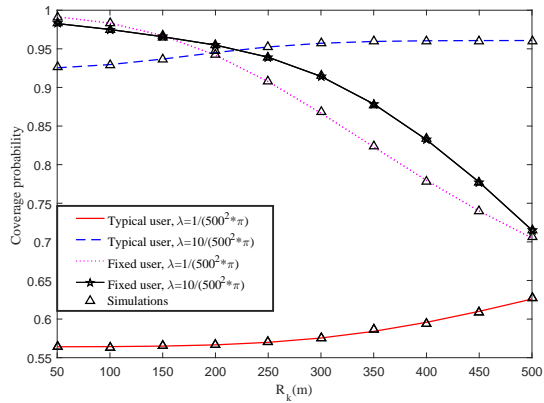}
\caption{Coverage probability of user-centric NOMA versus the distance of fixed users, with target rate $R_{t}=1$ BPCU and $R_f=0.5$ BPCU. The path loss exponent is fixed to $\alpha=3$, and the power of UAV is fixed to -30dBm.}
\label{UserCentric_diffRk}
\vspace{-0.3in}
\end{figure*}

In Fig.~\ref{UserCentric_diffRk}, the impact of different choices of UAV density and the distance of fixed users is studied. As can be observed from the figure, increasing the distance of fixed users will decrease the coverage probability for fixed users, whereas the coverage probability of typical users increases. This is due to the fact that the distance of fixed users has affect on user association for typical users. For fixed users, the received power decreases dramatically when the distance increases. On the other hand, for the dashed curve and star curve, where the density of UAV is 10 times greater than the solid curve and dotted curves, the coverage probability of typical NOMA users in the case of high UAV density is much greater than the case of low UAV density. This is because that the number of UAVs is increased, which leads to the decrease of the distance of connected UAV.
It is also worth noting that there are two crosses of fixed users, which mean that there exists an optimal distance of fixed users for the given UAV density.

\begin{figure*}[t!]
\centering
\includegraphics[width =2.8in]{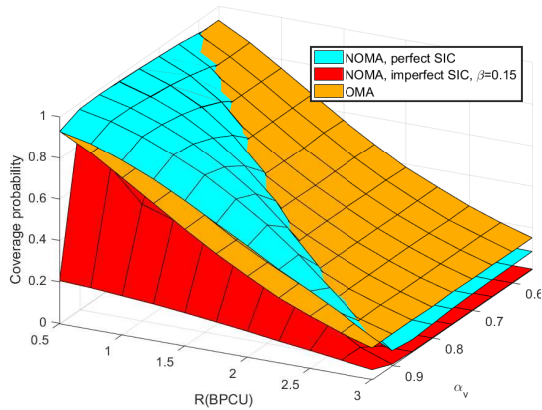}
\caption{Coverage probability of typical users versus targeted rate $R_{t}=R$ BPCU, and power allocation factor $\alpha_v$, with the imperfect SIC coefficient $\beta=0, 0.15$. The target rate of fixed users $R_f=0.5$ BPCU, and the horizontal distance of the fixed user is 300m. The transmit power of UAVs is fixed to -30dBm with path loss exponent $\alpha=3$. The fading parameters $m=3$ and $m_I=2$.}
\label{UserCentric_3D}
\vspace{-0.3in}
\end{figure*}

Next, Fig.~\ref{UserCentric_3D} plots the coverage probability of paired NOMA users in the user-centric strategy versus target rate $R$ and power allocation factor $\alpha_v$. It is observed that the coverage probability is zero in the case of inappropriate target rates and power allocation factors, which verifies the insights from {\bf Remark~\ref{user-centric power allocation}}.
The coverage probability of typical users in OMA is also plotted, which indicates that NOMA is capable for outperforming OMA for the appropriate power allocation factors and target rates of paired users.
One can also observe that NOMA cannot outperform OMA in the case of $\beta=0.15$ for the user-centric strategy. This indicates that hybrid NOMA/OMA assisted UAV framework may be a good solution in the case of poor SIC quality. The UAV could intelligently choose the access techniques for improving the system coverage probability.
\subsection{UAV-centric strategy}

\begin{figure*}[t!]
\centering
\subfigure[Coverage probability of the UAV-centric NOMA versus the transmit power in NLoS scenario, where the fading parameters $m=1$, $m_I=1$.]{\label{Outage_UAVCentric with m=1}
\includegraphics[width =2.8in]{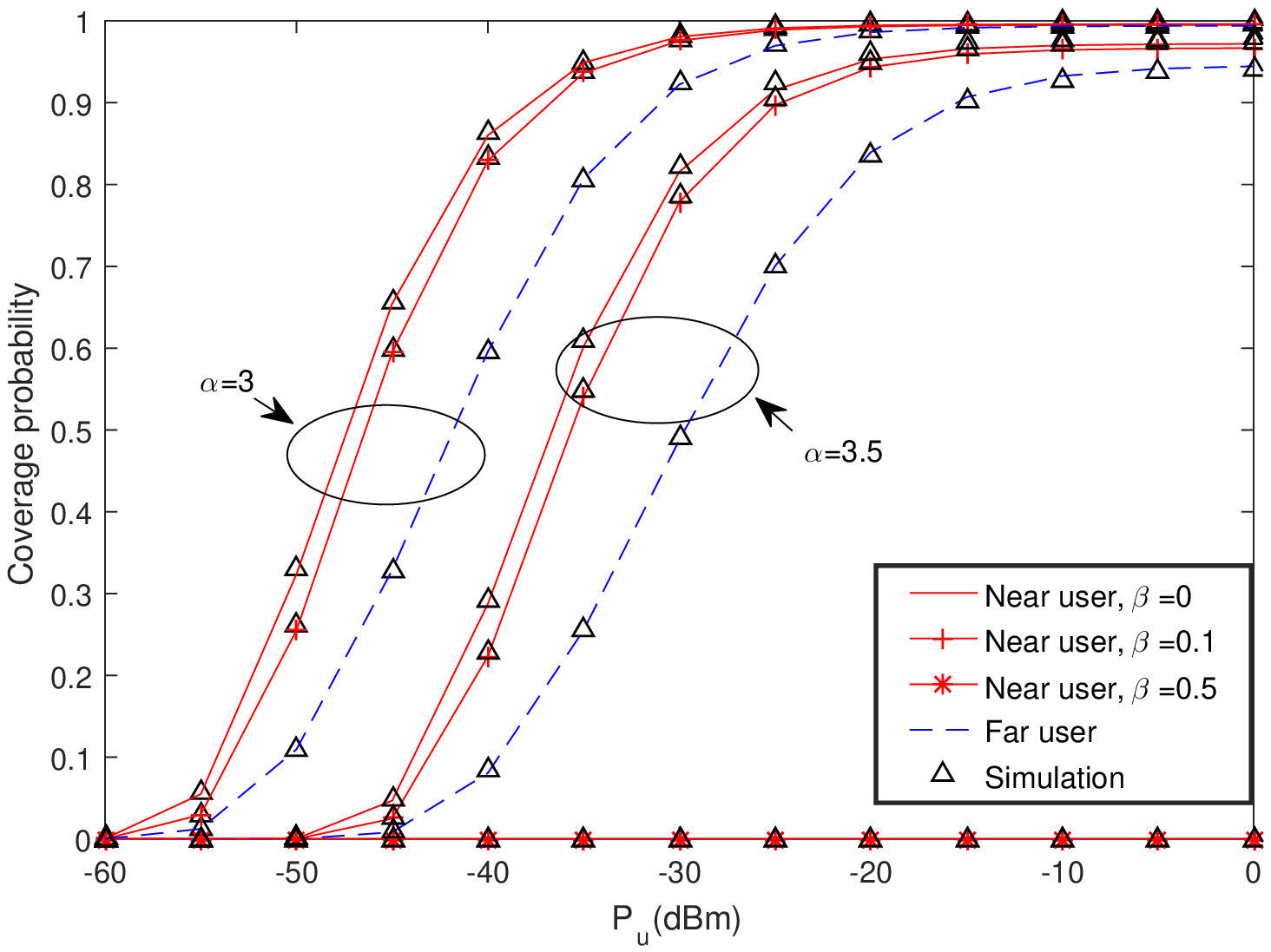}}
\subfigure[Coverage probability of the UAV-centric NOMA versus the transmit power in both NLoS and LoS scenarios with path loss exponent $\alpha=3.5$, where the fading parameters $m=2$, $m_I=1$.]{\label{Outage_UAVCentric with m=2}
\includegraphics[width= 2.8in]{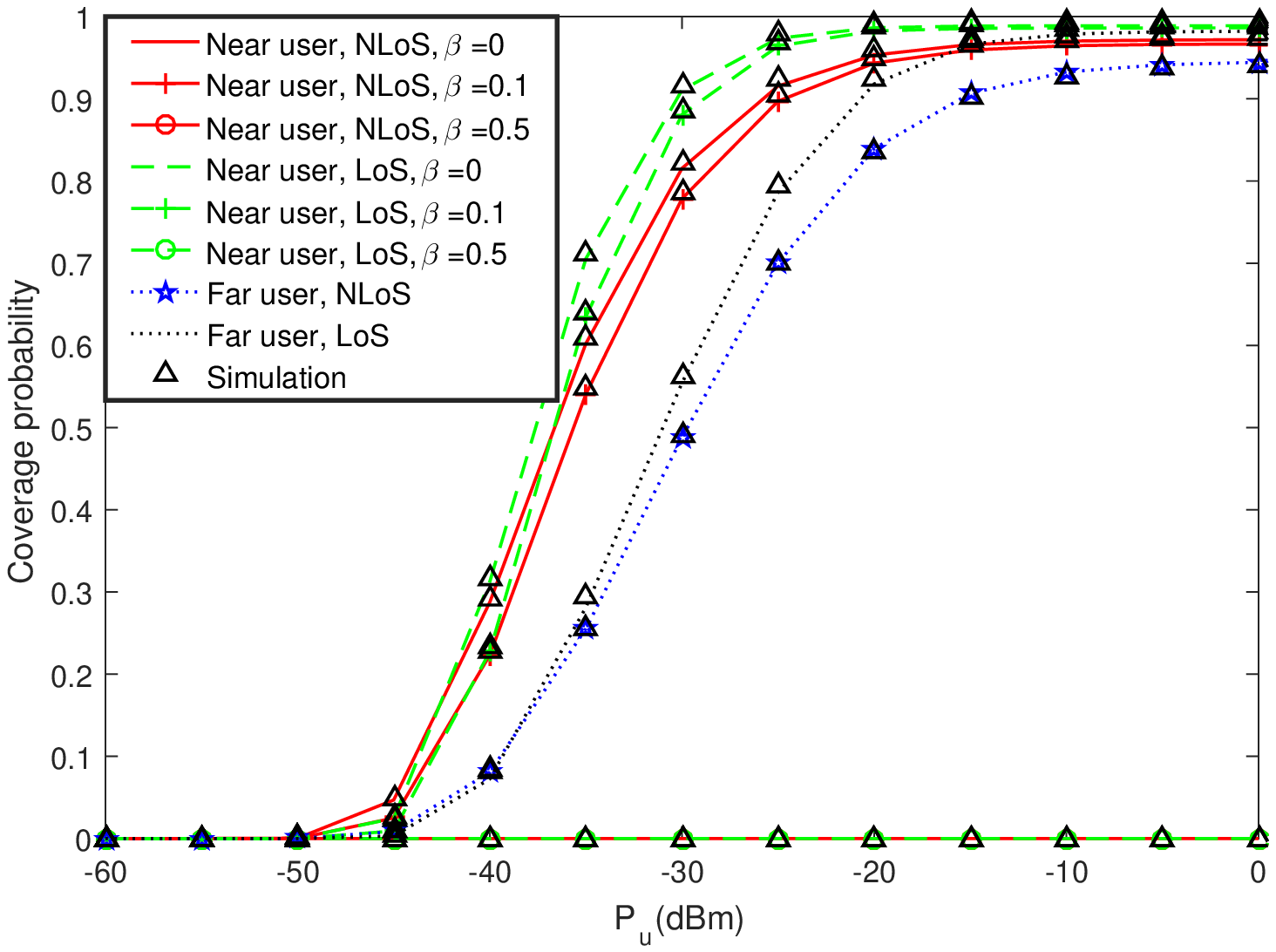}}
\caption{Coverage probability of paired NOMA users versus the transmit power, with target rate $R_{w}=1.5$ and $R_v=1$ BPCU, respectively. The exact results of NOMA are calculated from \eqref{coverage of near user expression}.}
\label{Fig5:Outage_UAVCetric}
\vspace{-0.3in}
\end{figure*}

In the UAV-centric strategy, $\varepsilon=0.1$m to evaluate the interference received from the UAV located at the distance $R$.
Then, we evaluate the performance of the downlink users in the UAV-centric strategy. In Figs.~\ref{Outage_UAVCentric with m=1} and~\ref{Outage_UAVCentric with m=2}, the impact of the NOMA assisted UAV-centric strategy in terms of the coverage probability is studied. The target rates of near users and far users are set as $R_{w}=1.5$ BPCU and $R_v=1$ BPCU, respectively. Solid curves and dashed curve are the coverage probability of near users and far users, respectively. An interesting phenomenon occurs in the UAV-centric strategy that in the case $\beta=0.5$, the coverage probability of near users is all zero, which indicates that the transmission is failed. This is again due to the fact that $\alpha_w^2- \beta\alpha_v^2\varepsilon_w<0$, which verifies our obtained insights in {\bf{Remark \ref{user-centric power allocation}}}.

Comparing Fig. \ref{Outage_UAVCentric with m=1} with Fig. \ref{Outage_UAVCentric with m=2}, one can observe that the impact of fading parameter $m$ on the coverage probability is also significant, which is due to the fact that the received power level is greater in the case of larger $m$. Again, we can see that the coverage probability is also one of near users in the case of $\beta=0.5$, which indicates that the LoS propagation has no effect on {\bf{Remark \ref{user-centric power allocation}}}. It is also worth noting that the coverage probability of the user-centric strategy is much greater than the UAV-centric strategy in the case of $\beta=0.5$, which indicates that the UAV-centric strategy is more susceptible to ipSIC factor than the user-centric strategy.

\begin{figure*}[t!]
\centering
\includegraphics[width =2.8in]{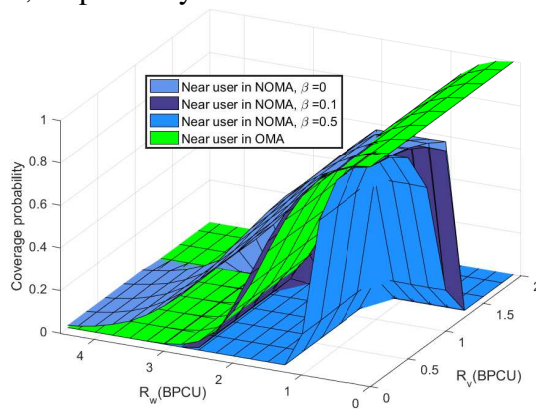}
\caption{Coverage probability of the near user versus the target rate. The transmit power of UAVs is fixed to -40dBm. The fading parameters $m=3$ and $m_I=2$.}
\label{UAV_3D}
\vspace{-0.3in}
\end{figure*}

Fig.~\ref{UAV_3D} plots the coverage probability for near users in the UAV-centric strategy in the cases of $\beta=0$, $\beta=0.1$, and $\beta=0.5$. One can obtain that on the one hand, inappropriate power allocation will lead to the coverage probability always being zero, which also verified {\bf Remark~\ref{user-centric power allocation}}. On the other hand, we can see that in the case of $\beta>0$, the coverage probability decreases dramatically when increasing target rate, which verified that the SIC residue is the dominant interference in NOMA. In order to provide more insights, the coverage performance of OMA in the UAV-centric strategy is also provided. We can see that in the case of $\beta=0$, NOMA performs better than OMA, which indicates that the proposed frameworks are analytically shown to be applicable for UAV communications. We can also see that in the case of~$\beta=0.1$, the coverage performance of NOMA and OMA assisted UAV cellular networks show closed agreement, which also indicates that hybrid NOMA/OMA assisted UAV framework may be a good solution for the UAV-centric strategy.

\section{Conclusions}
In this article, we first proposed an overview on a pair of important new paradigms in UAV assisted cellular communications, namely, user-centric strategy and UAV-centric strategy.
The user-centric strategy is applicable in the case when all the users located in the Voronoi cell are needed to be served by the UAV simultaneously. The derived results provide the benchmark for the NOMA assisted UAV cellular networks. The UAV-centric strategy is motivated by the fact that, in practice, it is more applicable to serve users in the dense networks. The key idea of the UAV-centric strategy is to provide services for the hotspot areas only, i.e., airports or resorts.
Then, the performance of proposed framework were evaluated, where multiple UAVs are distributed in the sky to serve multiple users on the terrestrial. Additionally, new analytical expressions for interference and coverage probability were derived for characterizing the performance in NOMA assisted UAV cellular frameworks.
An important future direction is to extend the 3-D distribution of interference sources to include other interfering UAVs located on the different heights.

\numberwithin{equation}{section}
\section*{Appendix~A: Proof of Lemma~\ref{lemma2:Lapalace transform of interference of typical user}} \label{Appendix:A}
\renewcommand{\theequation}{A.\arabic{equation}}
\setcounter{equation}{0}

Consider a HPPP $\Psi$ with density $\lambda$, the Laplace transform of the interference for the typical user can be expressed as follows:
\begin{equation}\label{laplace transform of typical user_user_centric in appendix}
\begin{aligned}
\mathcal{L}_t \left( {s} \right) &= {{\mathbb E}}\left\{ {\exp \left( { - s{I_{t,\Psi} }} \right)} \right\}
={\mathbb E}\left\{  { \exp  \left( { - s\sum\limits_{j \in \Psi,{d_j} > {r_{t}} } {{{\left| {{g_j}} \right|}^2}{\frac{{{P_u}}}{{{m_I}}}}d_{j}^{ - \alpha_I }} } \right) }   \right\}.
\end{aligned}
\end{equation}

Using the moment generating function (MGF) of Gamma random variable ${\left| {{g_j}} \right|}$, the Laplace transform can be rewritten to
\begin{equation}\label{laplace transform of typical user in appendix I2}
\begin{aligned}
\mathcal{L}_t \left( {s} \right) = \exp \left( { - 2\pi {\lambda}\int\limits_{{{r_{t}} }}^\infty  {\left( {1 - {{{\mathbb E}}_g}\left\{ {\exp \left( { - s{{\left| {{g_j}} \right|}^2}{\frac{{{P_u}}}{{{m_I}}}}r^{ - \alpha_I }} \right)} \right\}} \right)rdr} } \right).
\end{aligned}
\end{equation}

With the aid of Laplace transform for the Nakagami-$m$ distribution with fading parameter $m_I$, we can obtain ${{{\mathbb E}}_g}\left\{ {s{{\left| {{g_j}} \right|}^2}{{\frac{{{P_u}}}{{{m_I}}}}}r^{ - \alpha_I}} \right\} = {\left( {1 + \frac{{s{P_u}{r^{ - \alpha_I }}}}{m_I}} \right)^{ - m_I}}$. As such, by applying binomial expansion, the Laplace transform of the interference at the typical user can be rewritten to
\begin{equation}\label{laplace transform of typical user in appendix bionomial after}
\begin{aligned}
\mathcal{L}_t \left( {s} \right) &= \exp \left( { - 2\pi {\lambda}\int\limits_{{{r_{t}}}}^\infty  {\left( {1 - {{\left( {1 + \frac{{s{P_u}{r^{ - \alpha_I }}}}{{m_I}}} \right)}^{ - {{m_I}}}}} \right)}rdr } \right) \\
& = \exp \left( { - 2\pi {\lambda}\int\limits_{{r_{t}}}^\infty  {{\frac{{\sum\limits_{i{\rm{ = 0}}}^{{m_I}} { {
   {m_I}  \choose
   i}} {{\left( {\frac{{s{P_u}}}{{{{r^{\alpha_I} }}{m_I}}}} \right)}^i} - 1}}{{{\left( {1 + \frac{{s{P_u}{r^{ - \alpha_I }}}}{{{m_I}}}} \right)}^{{m_I}}}}} } rdr} \right).
\end{aligned}
\end{equation}

Then, after some algebraic manipulations, we have
\begin{equation}\label{Laplace transform of the typical expression after bonomial in appendix}
\begin{aligned}
\mathcal{L}_t \left( {s} \right) &= \exp \left( { - 2\pi {\lambda}\sum\limits_{i{\rm{ = 1}}}^{m_I} { {
   {m_I}  \choose
   i} } {{\left( {\frac{{s{P_u}}}{{m_I}}} \right)}^i}  \int\limits_{{{r_{t}}}}^\infty  {\frac{{{r^{ - \alpha_I i + 1}}}}{{{{\left( {1 + \frac{{s{P_u}{r^{ - \alpha_I }}}}{{m_I}}} \right)}^{m_I}}}}dr} } \right)\\
&\mathop {(a)}\limits_ =  \exp \left( { - \frac{{2\pi {\lambda}}}{\alpha_I }\sum\limits_{i{\rm{ = 1}}}^{{m_I}} { {
   {m_I}  \choose
   i} } {{\left( {\frac{{s{P_u}}}{{{m_I}}}} \right)}^{\delta_I} }{{\left( { - 1} \right)}^{\delta_I  - 1}}\int\limits_0^{ - \frac{{s{P_u}}}{{r_{t}^{\alpha_I} {m_I}}}} {\frac{{{t^{i - \delta_I  - 1}}}}{{{{\left( {1 - t} \right)}^{{m_I}}}}}dt} } \right),
\end{aligned}
\end{equation}
where $(a)$ is obtained by using $t={ - \frac{{s{P_u}}}{{r^{\alpha_I} {m_I}}}}$. Based on \cite[eq. (8.391)]{Table_of_integrals}, we can finally obtain the Laplace transform of the interference in the user-centric strategy in~\eqref{laplace transform of typical user in lemma}.

\numberwithin{equation}{section}
\section*{Appendix~B: Proof of Lemma~\ref{lemma3:outage of the typical user near user case}} \label{Appendix:B}
\renewcommand{\theequation}{B.\arabic{equation}}
\setcounter{equation}{0}

Then, we derive the coverage probability of the typical user as
\begin{equation}\label{outage first expression appendix C}
\begin{aligned}
&{P_{t,near}}\left( r \right) = {{\mathbb E}_{{I_\Psi }}}\left\{ {{\rm{Pr}} \left( {{{\left| {{h_w}} \right|}^2} < {M_{t*}}\left( {{\sigma ^2} + {I_\Psi }} \right)r_t^\alpha } \right)} \right\}\\
& = \exp \left( { - m{M_{t*}}{\sigma ^2}r_t^\alpha } \right){{\mathbb E}_{{I_{\Psi }}}}\left\{ {\exp \left( { - m{M_{t*}}{I_\Psi }r_t^\alpha } \right)} \right\}\sum\limits_{n = 0}^{m - 1} {\frac{{{{\left( {m{M_{t*}}\left( {{\sigma ^2} + {I_\Psi }} \right)r_t^\alpha } \right)}^n}}}{{n!}}}  \\
&= \sum\limits_{n = 0}^{m - 1} {\sum\limits_{p = 0}^n { {
   {n}  \choose
   p}} \frac{{r_t^{\alpha n}}{{{( - 1)}^n}}}{{n!}}} \underbrace {\exp \left( { - m{M_{t*}}{\sigma ^2}r_t^\alpha } \right){{\left( {-m{M_{t*}}{\sigma ^2}} \right)}^p}}_{{\Lambda _1}}\underbrace {{{\mathbb E}_{{I_{\Psi }}}}\left\{ {\exp \left( { - m{M_{t*}}{I_\Psi }r_t^\alpha } \right)} \right\}{{\left( {-m{M_t}{I_\Psi }} \right)}^{n - p}}}_{{\Lambda _2}}.
\end{aligned}
\end{equation}

Using the fact that
\begin{equation}\label{faadi prepare}
{\left. {\frac{{{d^p}\left( {\exp \left( { - m{M_{t*}}{\sigma ^2}y} \right)} \right)}}{{d{y^p}}}} \right|_{y = r_t^\alpha }} = \exp \left( { - m{M_{t*}}{\sigma ^2}r_t^\alpha } \right){\left( {-m{M_{t*}}{\sigma ^2}} \right)^p},
\end{equation}
we can have
\begin{equation}\label{faadi expression}
{\Lambda _1}{\rm{ = }} {\left. {\frac{{{d^p}\left( {\exp \left( { - m{M_{t*}}{\sigma ^2}y} \right)} \right)}}{{d{y^p}}}} \right|_{y = r_t^\alpha }}.
\end{equation}

Now, we apply the Fa \`{a} di Bruno's formula to solve the derivative of $p$-th order as follows:
\begin{equation}\label{faadi fomular for 1 in appendix}
{\Lambda _1} = \exp \left( { - m{M_{t*}}{\sigma ^2}r_t^\alpha } \right)\sum {p!} \prod\limits_{j = 1}^p {\frac{{{{\left( {\left( { - m{M_{t*}}{\sigma ^2}} \right)\prod\limits_{k = 0}^{j-1} {\left( {1 - k} \right)r_t^{\alpha (1 - j)}} } \right)}^{{q_j}}}}}{{{q_j}!{{\left( {j!} \right)}^{{q_j}}}}}},
\end{equation}
where the sum ${q_j}$ is over all p-tuples of nonnegative integers satisfying the constraint
\begin{equation}\label{qj constraint}
1 \cdot {q_1} + 2 \cdot {q_2} +  \cdots  + p \cdot {q_p} = p.
\end{equation}

Similar to the steps from \eqref{faadi prepare} to \eqref{faadi fomular for 1 in appendix}, ${\Lambda _2}$ can be expressed to
\begin{equation}\label{lamda 2 faadi prepare}
\begin{aligned}
&{\Lambda _2}{\rm{ = }}{{\mathbb E}_{{I_{\Psi }}}}\left\{ {\exp \left( { - m{M_{t*}}{I_\Psi }r_t^\alpha } \right){{\left( {-m{M_{t*}}{I_\Psi }} \right)}^{n - p}}} \right\}\\
&  = {{\mathbb E}_{{I_{\Psi }}}}\left\{ {{{\left. {\frac{{{d^{n - p}}\left( {\exp \left( { - m{M_{t*}}{I_\Psi }y} \right)} \right)}}{{d{y^{n - p}}}}} \right|}_{y = r_t^\alpha }}} \right\}\\
& = {\left. {\frac{{{d^{n - p}}\left( {m{\mathcal{L}_t}\left( {{M_{t*}}y } \right)} \right)}}{{d{y^{n - p}}}}} \right|_{y = r_t^\alpha }}.
\end{aligned}
\end{equation}

It is challenging to derive ($n-p$)-th order derivation of incomplete Beta function directly. Thus, the derived incomplete Beta function in \eqref{laplace transform of typical user in lemma} can be written to
\begin{equation}\label{beta function expansiong}
\begin{aligned}
B\left( {\frac{{-s{P_u}}}{{{m_I}r_{t}^{\alpha_I} }};i - \delta_I ,1 - {m_I}} \right) &=  {\left( {\frac{{-s{P_u}}}{{{m_I}r_{t}^{\alpha_I} }}} \right)^{\left( {i - \delta_I} \right)}}\frac{1}{{i - \delta_I}}{}_2{F_1}\left( {1 - \delta_I,{m_I};2 - \delta_I;\left( {\frac{{-s{P_u}}}{{{m_I}r_{t}^{\alpha_I} }}} \right)} \right) \\
& = {\left( {\frac{{-s{P_u}}}{{{m_I}r_{t}^{\alpha_I} }}} \right)^{\left( {i - \delta_I + a} \right)}}\sum\limits_{a = 0}^\infty  {\frac{{{{\left( {{m_I}} \right)}_a}}}{{a!\left( {i - \delta_I + a} \right)}}},
\end{aligned}
\end{equation}
where ${\left( {{m_I}} \right)_a}$ denotes rising Pochhammer symbol, which can be calculated as $\frac{{\Gamma \left( {{m_I} + a} \right)}}{{\Gamma \left( {{m_I}} \right)}}$.

Thus, the Laplace transform can be rewritten to
\begin{equation}\label{lambda 2 before faadi}
\begin{aligned}
{{\cal L}_t}\left( s \right) = \exp \left( { - \frac{{2\pi \lambda }}{\alpha_I }\sum\limits_{a = 0}^\infty  {\frac{{{{\left( {{m_I}} \right)}_a}}}{{a!\left( {i - \delta_I  + a} \right)}}} \sum\limits_{i{\rm{ = 1}}}^{{m_I}} { {
   {m_I}  \choose
   i} } {{\left( {\frac{{s{P_u}}}{{{m_I}}}} \right)}^{i + a}}{{\left( { - 1} \right)}^a}r_t^{ - \alpha_I \left( {i - \delta_I  + a} \right)}} \right).
\end{aligned}
\end{equation}

Then, substituting \eqref{lambda 2 before faadi} into \eqref{lamda 2 faadi prepare}, and using Fa \`{a} di Bruno's formula, \eqref{lamda 2 faadi prepare} can be transformed into
\begin{equation}\label{faadi trans lambda 2}
\begin{aligned}
&{\Lambda _2}={\left. {\frac{{{d^{n - p}}\left( {\exp \left( { - {\Lambda _3}{y^{\delta}}} \right)} \right)}}{{d{y^{n - p}}}}} \right|_{y = r_t^\alpha }}\\
&=\exp \left( { - {\Lambda _3}r_t^{2+ (\alpha-\alpha_I)(i+a)   }} \right)\sum {(n - p)!} \prod\limits_{b = 1}^{n - p} {\frac{{{{\left( {\left( { - {\Lambda _3}} \right)\prod\limits_{k = 0}^{b - 1} {\left( {\delta - k} \right)r_t^{2 +(\alpha-\alpha_I)(i+a) - \alpha b}} } \right)}^{{q_b}}}}}{{{q_b}!{{\left( {b!} \right)}^{{q_b}}}}}},
\end{aligned}
\end{equation}
where ${\Lambda _3} = \frac{{2\pi \lambda }}{\alpha_I }\sum\limits_{a = 0}^\infty  {\frac{{{{\left( {{m_I}} \right)}_a}}}{{a!\left( {i - \delta_I  + a} \right)}}} \sum\limits_{i{\rm{ = 1}}}^{{m_I}} {{
   {m_I}  \choose
   i} } {\left( {\frac{{{M_{t*}}{P_u}}}{{{m_I}}}} \right)^{i + a}}{\left( { - 1} \right)^a}$, and ${q_b}$ is over all $(n-p)$-tuples of nonnegative integers satisfying the constraint $1 \cdot {q_1} + 2 \cdot {q_2} +  \cdots  + (n - p) \cdot {q_b} = (n - p)$.

Substituting \eqref{faadi fomular for 1 in appendix} and \eqref{faadi trans lambda 2} into \eqref{outage first expression appendix C}, we can derive the coverage probability conditioned on the distance for the typical user in the near user case, as given in \eqref{coverage probability typical user near Lemma}. The proof is complete.

\numberwithin{equation}{section}
\section*{Appendix~C: Proof of Lemma~\ref{Lapalace transform of interference of near user}} \label{Appendix:C}
\renewcommand{\theequation}{C.\arabic{equation}}
\setcounter{equation}{0}

Unlike the user-centric strategy, the interfering UAV located at the distance $R$ is necessary to evaluate separately in the UAV-centric strategy.
In this section, we evaluate the Laplace transform of inter-cell interference of the near user in the UAV-centric strategy, where the inter-cell interference experience at the near user can be given by
\begin{equation}\label{intercell interference}
{I_{w,\Psi} } = \underbrace {\sum\limits_{j \in \Psi ,{d_j} > R} {{{\left| {{g_j}} \right|}^2}{P_u}} d_j^{ - \alpha_I }}_{{I_2}} + \underbrace {{{\left| {{g_1}} \right|}^2}{P_u}{R^{ - \alpha_I }}}_{{I_1}},
\end{equation}
where ${I_1}$ denotes the received power from the interfering UAV located at the distance $R$, and ${I_2}$ denotes the received power from all other interfering UAVs except the one located at the distance $R$.
For the near user at the typical cell, the Laplace transform of interference power distribution conditioned on the serving distance $R$ is given by
\begin{equation}\label{laplace transform of near user in appendix}
\begin{aligned}
\mathcal{L}_w \left( {s\left| R \right.} \right) &= {{\mathbb E}}\left\{ {\exp \left( { - s{I_{w,\Psi} }} \right)\left| R \right.} \right\}
={\mathbb E}\left\{  {\left. \exp  \left( { - s\sum\limits_{j \in \Psi,{d_j} > R } {{{\left| {{g_j}} \right|}^2}{\frac{{{P_u}}}{{{m_I}}}}d_{j}^{ - \alpha_I }} }  - s{{{\left| {{g_1}} \right|}^2}{\frac{{{P_u}}}{{{m_I}}}}R^{ - \alpha_I }} \right) \right|} R  \right\} \\
&= {{ {{\mathbb E}}}}\left\{ {\exp \left. {\left( { - s\sum\limits_{j \in \Psi, d_j>R } {{{ {{\mathbb E}}}_g}\left\{ {{{\left| {{g_j}} \right|}^2}} \right\}{\frac{{{P_u}}}{{{m_I}}}}d_{j}^{ - \alpha_I }} }  - s{{{\mathbb E}_g}\left\{ {{{\left| {{g_1}} \right|}^2}} \right\}{\frac{{{P_u}}}{{{m_I}}}}R^{ - \alpha_I }} \right)} \right|R} \right\}.
\end{aligned}
\end{equation}

We first evaluate the Laplace transform of $I_2$. Using the MGF of Gamma random variable ${\left| {{g_j}} \right|}$, and after some algebraic manipulations, $I_2$ can be rewritten to
\begin{equation}\label{laplace transform of near user in appendix I2}
\begin{aligned}
I_2 = \exp \left( { - 2\pi {\lambda }\int\limits_{{l_I}}^\infty  {\left( {1 - {{{\mathbb E}}_g}\left\{ {\exp \left( { - s{{\left| {{g_j}} \right|}^2}{\frac{{{P_u}}}{{{m_I}}}}r^{ - \alpha_I }} \right)} \right\}} \right)rdr} } \right),
\end{aligned}
\end{equation}
where ${l_I} = \sqrt {{R ^2} + {h^2}}$.

Similar to the arguments from \eqref{laplace transform of typical user in appendix I2} to \eqref{Laplace transform of the typical expression after bonomial in appendix}, the Laplace transform of $I_2$ can be obtained as
\begin{equation}\label{laplace transform of near user in appendix I2}
\begin{aligned}
I_2 = \exp \left( { - \frac{{2\pi {\lambda }}}{\alpha_I }\sum\limits_{i{\rm{ = 1}}}^{m_I} {{
   {m_I}  \choose
   i}} {{\left( {\frac{{s{P_u}}}{{m_I}}} \right)}^{\delta_I}}(-1)^{(\delta_I-i)}B\left( {\frac{{-s{P_u}}}{{l_I^{\alpha_I }m_I}};i - \delta_I,1-{m_I}} \right)} \right).
\end{aligned}
\end{equation}

Note that the MGF derived in \eqref{laplace transform of near user in appendix I2} does not include the interfering UAV located at the distance $R$ strictly, which is actually the largest interference source. Therefore, using Poisson Hole Process (PHP), the first interference located at distance $R$ can be derived as follows:
\begin{equation}\label{first interferece_in appendix_I1}
I_1=\exp \left( {\frac{{ - 2\pi  }}{{\pi {{\left( {R + \varepsilon } \right)}^2} - \pi {{\left( {R - \varepsilon } \right)}^2}}}\int\limits_{{l_I} - \varepsilon }^{{l_I} + \varepsilon } {\left( {1 - {{{\mathbb E}}_g}\left\{ {\exp \left( { - s{{\left| {{g_1}} \right|}^2}\frac{{{P_u}}}{{{m_I}}}{{l_I^ {- \alpha_I }}}} \right)} \right\}} \right)rdr} } \right),
\end{equation}
where $\varepsilon$ is a small distance to evaluate the first interfering UAV.

With the aid of Laplace transform for the Nakagami-$m$ distribution with fading parameter $m_I$, we can obtain ${{{\mathbb E}}_g}\left\{ {{{\left| {{g_1}} \right|}^2}{{\frac{{{P_u}}}{{{m_I}}}}}d_1^{ - \alpha_I }} \right\} = {\left( {1 + \frac{{s{P_u}{l_I^{ - \alpha_I }}}}{m_I}} \right)^{ - m_I}}$. As such, the Laplace transform of the first interfering UAV can be rewritten to

\begin{equation}\label{Laplace transform of near expression I1 in appendix}
\begin{aligned}
I_1 &= \exp \left( {\frac{{ - 1}}{{2R\varepsilon }}\int\limits_{{l_I} - \varepsilon }^{{l_I} + \varepsilon } {\left( {1 - {{\left( {1 + \frac{{s{P_u}}}{{{m_I}{l_I^ { \alpha_I }}}}} \right)}^{ - {m_I}}}} \right)rdr} } \right)\\
& = \exp \left( { - \frac{{{l_I}}}{R} + \frac{{{l_I}}}{R}{{\left( {1 + \frac{{s{P_u}}}{{{l_I^{\alpha_I} }{m_I}}}} \right)}^{ - {m_I}}}} \right).
\end{aligned}
\end{equation}

Based on \eqref{laplace transform of near user in appendix I2} and \eqref{Laplace transform of near expression I1 in appendix}, we can obtain the Laplace transform of the near user for the UAV-Centric strategy as given in \eqref{laplace transform of UAV in lemma}. The proof is complete.

\numberwithin{equation}{section}
\section*{Appendix~D: Proof of Lemma~\ref{lemma5:outage of the near user conditioned on radius}} \label{Appendix:D}
\renewcommand{\theequation}{D.\arabic{equation}}
\setcounter{equation}{0}

In order to prove the desired result, and according to Newton's Generalization of the binomial theorem \cite{Mathematics}, we first transform \eqref{Laplace transform of near expression I1 in appendix} into
\begin{equation}\label{Newton's Generalization}
{I_1} = \exp \left( { - \frac{{{l_I}}}{R} + \frac{{{l_I}}}{R}\sum\limits_{U = 0}^\infty  {{{( - 1)}^U}C_{{m_I} + U + 1}^U} {{\left( {\frac{{s{P_u}}}{{l_I^\alpha {m_I}}}} \right)}^U}} \right),
\end{equation}
where $C_{{m_I} + U + 1}^U = \frac{{({m_I} + U + 1)({m_I} + U) +  \cdots ({m_I} + 2)}}{{k!}}$.

According to the SINR expressions in \eqref{SINR_w tov} and \eqref{SINR_w}, and similar to Appendix B, we can derive the coverage probability conditioned on the serving distance $R$ of the near user in the UAV-centric strategy to
\begin{equation}\label{UAV-Centric_cover_first of near in appendix}
\begin{aligned}
{P_{cov,w}}\left( {r\left| R \right.} \right) = \exp \left( { - m{M_{w*}}\left( {{\sigma ^2} + {I_1} + {I_2}} \right)r_w^\alpha } \right)\sum\limits_{n = 0}^{m - 1} {\frac{{{{\left( {m{M_{w*}}\left( {{\sigma ^2} + {I_1} + {I_2}} \right)r_w^\alpha } \right)}^n}}}{{n!}}} .
\end{aligned}
\end{equation}
By applying polynomial expansion to \eqref{UAV-Centric_cover_first of near in appendix}, the coverage probability can be rewritten to
\begin{equation}\label{UAV-Centric_cover_after polynomial in appendix}
\begin{aligned}
{P_{cov,w}}\left( {r\left| R \right.} \right)& = \sum\limits_{n = 0}^{m - 1} {\sum\limits_{k = 0}^n {\sum\limits_{l = 0}^k {\frac{{{{( - 1)}^n}r_w^{\alpha n}}}{{l!(k - l)!(n - k)!}}} } } \exp \left( { - m{M_{w*}}{\sigma ^2}r_w^\alpha } \right){\left( { - m{M_{w*}}{\sigma ^2}} \right)^{n - k}} \\
& \times {{\mathbb E}_{{I_1}}}\left\{ {\exp \left( { - m{M_{w*}}{I_1}r_w^\alpha } \right){{\left( { - m{M_{w*}}{I_1}} \right)}^{k - l}}} \right\}{{\mathbb E}_{{I_2}}}\left\{ {\exp \left( { - m{M_{w*}}{I_2}r_w^\alpha } \right){{\left( { - m{M_{w*}}{I_2}} \right)}^l}} \right\}.
\end{aligned}
\end{equation}

Following the similar steps from \eqref{outage first expression appendix C} to \eqref{faadi trans lambda 2}, and according to Fa \`{a} di Bruno's formula,
we can readily derive that the first interference $I_1$ to
\begin{equation}\label{UAV-Centric I1 appendix last}
\begin{aligned}
{\left. {\frac{{{d^{k - l}}  {\mathcal{L}} ({M_{w*}}x)}}{{d{x^{k - l}}}}} \right|_{x = r_w^\alpha }} = \exp \left( { - \frac{{{l_I}}m}{R} + {\Theta _2}r_w^{\alpha U}} \right)\sum {(k - l)!} \prod\limits_{j = 1}^{k - l} {\frac{{{{\left( {\left( { - {\Theta _2}} \right)\prod\limits_{p = 0}^{j - 1} {\left( {u - p} \right)r_w^{\alpha (U - j)}} } \right)}^{{q_u}}}}}{{{q_u}!{{\left( {j!} \right)}^{{q_u}}}}}},
\end{aligned}
\end{equation}
where ${\Theta _2} = \frac{{{l_I}}m}{R}\sum\limits_{U = 0}^\infty  {{{( - 1)}^U}C_{{m_I} + U + 1}^u} {\left( {\frac{{{M_{w*}}{P_u}}}{{l_I^{\alpha_I} {m_I}}}} \right)^U}$.
Then, the closed-form expression of the coverage probability for the near user in \eqref{coverage probability conditioned on radius Lemma near user UAV-centric} can be obtained. Thus, the Lemma is proved.

\begin{spacing}{1}
\bibliographystyle{IEEEtran}%
\bibliography{IEEEabrv,bib2018}

\begin{thebibliography}{10}
\providecommand{\url}[1]{#1}
\csname url@samestyle\endcsname
\providecommand{\newblock}{\relax}
\providecommand{\bibinfo}[2]{#2}
\providecommand{\BIBentrySTDinterwordspacing}{\spaceskip=0pt\relax}
\providecommand{\BIBentryALTinterwordstretchfactor}{4}
\providecommand{\BIBentryALTinterwordspacing}{\spaceskip=\fontdimen2\font plus
\BIBentryALTinterwordstretchfactor\fontdimen3\font minus
  \fontdimen4\font\relax}
\providecommand{\BIBforeignlanguage}[2]{{%
\expandafter\ifx\csname l@#1\endcsname\relax
\typeout{** WARNING: IEEEtran.bst: No hyphenation pattern has been}%
\typeout{** loaded for the language `#1'. Using the pattern for}%
\typeout{** the default language instead.}%
\else
\language=\csname l@#1\endcsname
\fi
#2}}
\providecommand{\BIBdecl}{\relax}
\BIBdecl

\bibitem{multi_VTC2019}
T.~Hou, Y.~Liu, Z.~Song, X.~Sun, and Y.~Chen, ``Non-orthogonal multiple access
  in multi-{UAV} networks,'' in \emph{The 2019 IEEE 90th Vehicular Technology
  Conference}, Hawaii, USA, Sep. 2019, pp. 1--1.

\bibitem{UAV_zeng}
Y.~Zeng, R.~Zhang, and T.~J. Lim, ``Wireless communications with unmanned
  aerial vehicles: opportunities and challenges,'' \emph{IEEE Commun. Mag.},
  vol.~54, no.~5, pp. 36--42, May 2016.

\bibitem{Saad_D2D_UAV}
M.~Mozaffari, W.~Saad, M.~Bennis, and M.~Debbah, ``Unmanned aerial vehicle with
  underlaid device-to-device communications: Performance and tradeoffs,''
  \emph{IEEE Trans. Wireless Commun.}, vol.~15, no.~6, pp. 3949--3963, Jun.
  2016.

\bibitem{3GPP_36.777}
``Study on enhanced {LTE} support for aerial vehicles (release 15),''
  \emph{Online: ftp://www.3gpp.org/specs/archive/36\_series/36.777}, vol. 3GPP
  TR 36.777, Jun. 2017.

\bibitem{NOMA_mag_Ding}
Z.~Ding, Y.~Liu, J.~Choi, Q.~Sun, M.~Elkashlan, C.~I, and H.~V. Poor,
  ``Application of non-orthogonal multiple access in {LTE} and 5{G} networks,''
  \emph{IEEE Commun. Mag.}, vol.~55, no.~2, pp. 185--191, Feb. 2017.

\bibitem{wireless_sparse}
Z.~Qin, J.~Fan, Y.~Liu, Y.~Gao, and G.~Y. Li, ``Sparse representation for
  wireless communications: A compressive sensing approach,'' \emph{IEEE Signal
  Process. Mag.}, vol.~35, no.~3, pp. 40--58, May 2018.

\bibitem{PairingDING2016}
Z.~Ding, P.~Fan, and H.~V. Poor, ``Impact of user pairing on 5{G} nonorthogonal
  multiple-access downlink transmissions,'' \emph{IEEE Trans. Veh. Technol.},
  vol.~65, no.~8, pp. 6010--6023, Aug. 2016.

\bibitem{Massive_NOMA_Cellular_IoT}
M.~{Shirvanimoghaddam}, M.~{Dohler}, and S.~J. {Johnson}, ``Massive
  non-orthogonal multiple access for cellular {IoT}: Potentials and
  limitations,'' \emph{IEEE Commun. Mag.}, vol.~55, no.~9, pp. 55--61, Sep.
  2017.

\bibitem{NOMA_5G_beyond_Liu}
Y.~Liu, Z.~Qin, M.~Elkashlan, Z.~Ding, A.~Nallanathan, and L.~Hanzo,
  ``Nonorthogonal multiple access for 5{G} and beyond,'' \emph{Proc. of the
  IEEE}, vol. 105, no.~12, pp. 2347--2381, Dec. 2017.

\bibitem{Resource_allo_Islam}
S.~M.~R. Islam, M.~Zeng, O.~A. Dobre, and K.~Kwak, ``Resource allocation for
  downlink {NOMA} systems: Key techniques and open issues,'' \emph{IEEE
  Wireless Commun.}, vol.~25, no.~2, pp. 40--47, Apr. 2018.

\bibitem{Islam_NOMA_survey}
S.~M.~R. Islam, N.~Avazov, O.~A. Dobre, and K.~Kwak, ``Power-domain
  non-orthogonal multiple access ({NOMA}) in 5{G} systems: Potentials and
  challenges,'' \emph{IEEE Commun. Surveys Tuts.}, vol.~19, no.~2, pp.
  721--742, Secondquarter 2017.

\bibitem{heterNOMA_Qin}
Z.~Qin, X.~Yue, Y.~Liu, Z.~Ding, and A.~Nallanathan, ``User association and
  resource allocation in unified non-orthogonal multiple access enabled
  heterogeneous ultra dense networks,'' \emph{IEEE Commun. Mag.}, vol.~56,
  no.~6, pp. 86--92, Jun. 2018.

\bibitem{UAV_Channel}
A.~A. Khuwaja, Y.~Chen, N.~Zhao, M.~Alouini, and P.~Dobbins, ``A survey of
  channel modeling for {UAV} communications,'' \emph{IEEE Commun. Surveys
  Tuts.}, vol.~20, no.~4, pp. 2804--2821, Fourthquarter 2018.

\bibitem{A2A_UAV_Rice}
N.~Goddemeier and C.~Wietfeld, ``Investigation of air-to-air channel
  characteristics and a {UAV} specific extension to the {R}ice model,'' in
  \emph{2015 IEEE Globecom Workshops (GC Wkshps)}, Dec. 2015, pp. 1--5.

\bibitem{Rayleigh_UAV}
F.~Jiang and A.~L. Swindlehurst, ``Optimization of {UAV} heading for the
  ground-to-air uplink,'' \emph{IEEE J. Sel. Areas Commun.}, vol.~30, no.~5,
  pp. 993--1005, Jun. 2012.

\bibitem{UAV_finite_downlink}
V.~V. Chetlur and H.~S. Dhillon, ``Downlink coverage analysis for a finite
  3-{D} wireless network of unmanned aerial vehicles,'' \emph{IEEE Trans.
  Commun.}, vol.~65, no.~10, pp. 4543--4558, Oct. 2017.

\bibitem{wireless_communication_goldsmith}
A.~Goldsmith, \emph{Wireless Communication}.\hskip 1em plus 0.5em minus
  0.4em\relax Cambridge University Press, 2nd ed, 2010.

\bibitem{UAV_Trajectory_shuowen}
S.~Zhang, Y.~Zeng, and R.~Zhang, ``Cellular-enabled {UAV} communication: A
  connectivity-constrained trajectory optimization perspective,'' \emph{IEEE
  Trans. Commun.}, vol.~67, no.~3, pp. 2580--2604, Mar. 2019.

\bibitem{Lyu_UAV_hotspots}
J.~{Lyu}, Y.~{Zeng}, and R.~{Zhang}, ``{UAV}-aided offloading for cellular
  hotspot,'' \emph{IEEE Trans. Wireless Commun.}, vol.~17, no.~6, pp.
  3988--4001, Jun. 2018.

\bibitem{Dai_NOMA_survey}
L.~{Dai}, B.~{Wang}, Z.~{Ding}, Z.~{Wang}, S.~{Chen}, and L.~{Hanzo}, ``A
  survey of non-orthogonal multiple access for 5{G},'' \emph{IEEE Commun.
  Surveys Tuts.}, vol.~20, no.~3, pp. 2294--2323, thirdquarter 2018.

\bibitem{Shin_NOMA_cellular}
W.~{Shin}, M.~{Vaezi}, B.~{Lee}, D.~J. {Love}, J.~{Lee}, and H.~V. {Poor},
  ``Non-orthogonal multiple access in multi-cell networks: Theory, performance,
  and practical challenges,'' \emph{IEEE Commun. Mag.}, vol.~55, no.~10, pp.
  176--183, Oct. 2017.

\bibitem{Randomly_ding}
Z.~Ding, Z.~Yang, P.~Fan, and H.~V. Poor, ``On the performance of
  non-orthogonal multiple access in 5{G} systems with randomly deployed
  users,'' \emph{IEEE Signal Process. Lett.}, vol.~21, no.~12, pp. 1501--1505,
  Dec. 2014.

\bibitem{Liu_Coop_NOMA_SWIPT}
Y.~Liu, Z.~Ding, M.~Elkashlan, and H.~V. Poor, ``Cooperative non-orthogonal
  multiple access with simultaneous wireless information and power transfer,''
  \emph{IEEE J. Sel. Areas Commun.}, vol.~34, no.~4, pp. 938--953, Apr. 2016.

\bibitem{Yue_ISIC_2018}
X.~Yue, Y.~Liu, S.~Kang, A.~Nallanathan, and Y.~Chen, ``Modeling and analysis
  of two-way relay non-orthogonal multiple access systems,'' \emph{IEEE Trans.
  Commun.}, vol.~66, no.~9, pp. 3784--3796, Sep. 2018.

\bibitem{Nakagami_Hou}
T.~Hou, X.~Sun, and Z.~Song, ``Outage performance for non-orthogonal multiple
  access with fixed power allocation over {N}akagami-$m$ fading channels,''
  \emph{IEEE Commun. Lett.}, vol.~22, no.~4, pp. 744--747, Apr. 2018.

\bibitem{Clerckx_RSMA}
Y.~Mao, B.~Clerckx, and V.~O.~K. Li, ``Rate-splitting for downlink multi-user
  multi-antenna systems: Bridging {NOMA} and conventional linear precoding,''
  \emph{arXiv}, vol. 1710.11018v1, pp. 1--1, Aug. 2017.

\bibitem{Clerckx_RSMA_MIMO}
------, ``Rate-splitting for multi-antenna non-orthogonal unicast and multicast
  transmission: Spectral and energy efficiency analysis,'' \emph{arXiv}, vol.
  1808.08325v1, pp. 1--1, Aug. 2018.

\bibitem{Liu_physical_scurity_NOMA}
Y.~Liu, Z.~Qin, M.~Elkashlan, Y.~Gao, and L.~Hanzo, ``Enhancing the physical
  layer security of non-orthogonal multiple access in large-scale networks,''
  \emph{IEEE Trans. Wireless Commun.}, vol.~16, no.~3, pp. 1656--1672, Mar.
  2017.

\bibitem{energy_consumption_UAV}
Y.~{Zeng}, J.~{Xu}, and R.~{Zhang}, ``Energy minimization for wireless
  communication with rotary-wing {UAV},'' \emph{IEEE Trans. Wireless Commun.},
  vol.~18, no.~4, pp. 2329--2345, Apr. 2019.

\bibitem{Cellular-UAV_mag}
Y.~{Zeng}, J.~{Lyu}, and R.~{Zhang}, ``Cellular-connected {UAV}: Potential,
  challenges, and promising technologies,'' \emph{IEEE Wireless Commun.},
  vol.~26, no.~1, pp. 120--127, Feb. 2019.

\bibitem{UAV_general_Liu}
Y.~{Liu}, Z.~{Qin}, Y.~{Cai}, Y.~{Gao}, G.~Y. {Li}, and A.~{Nallanathan},
  ``{UAV} communications based on non-orthogonal multiple access,'' \emph{IEEE
  Wireless Commun.}, vol.~26, no.~1, pp. 52--57, Feb. 2019.

\bibitem{Liuxiao_Trajectory_multi-UAV}
X.~Liu, Y.~Liu, Y.~Chen, and L.~Hanzo, ``Trajectory design and power control
  for multi-{UAV} assisted wireless networks: A machine learning approach,''
  \emph{arXiv}, vol. 1812.07665v1, pp. 1--1, Dec. 2018.

\bibitem{UAV_NOMA_Trajectory}
N.~{Zhao}, X.~{Pang}, Z.~{Li}, Y.~{Chen}, F.~{Li}, Z.~{Ding}, and M.~{Alouini},
  ``Joint trajectory and precoding optimization for {UAV}-assisted {NOMA}
  networks,'' \emph{IEEE Trans. Commun.}, vol.~67, no.~5, pp. 3723--3735, May
  2019.

\bibitem{UAV_relay}
T.~M. Nguyen, W.~Ajib, and C.~Assi, ``A novel cooperative {NOMA} for designing
  {UAV}-assisted wireless backhaul networks,'' \emph{IEEE J. Sel. Areas
  Commun.}, vol.~36, no.~11, pp. 2497--2507, Nov. 2018.

\bibitem{Hou_Single_UAV}
T.~{Hou}, Y.~{Liu}, Z.~{Song}, X.~{Sun}, and Y.~{Chen}, ``Multiple antenna
  aided {NOMA} in {UAV} networks: A stochastic geometry approach,'' \emph{IEEE
  Trans. Commun.}, vol.~67, no.~2, pp. 1031--1044, Feb. 2019.

\bibitem{Uplink_NOMA_UAV}
W.~{Mei} and R.~{Zhang}, ``Uplink cooperative {NOMA} for cellular-connected
  {UAV},'' \emph{IEEE J. Sel. Areas Commun.}, vol.~13, no.~3, pp. 644--656,
  Jun. 2019.

\bibitem{UAV_multibeam_liangliu}
L.~Liu, S.~Zhang, and R.~Zhang, ``Exploiting {NOMA} for multi-beam {UAV}
  communication in cellular uplink,'' \emph{arXiv}, vol. 1810.10839v1, pp.
  1--1, Oct. 2018.

\bibitem{Han_millimeter_UAV}
K.~{Han}, K.~{Huang}, and R.~W. {Heath}, ``Connectivity and blockage effects in
  millimeter-wave air-to-everything networks,'' \emph{IEEE Wireless Commun.
  Lett.}, vol.~8, no.~2, pp. 388--391, Apr. 2019.

\bibitem{Satellite_UAV_network}
T.~Qi, W.~Feng, and Y.~Wang, ``Outage performance of non-orthogonal multiple
  access based unmanned aerial vehicles satellite networks,'' \emph{China
  Communications}, vol.~15, no.~5, pp. 1--8, May 2018.

\bibitem{UAV_emergency_disasters}
N.~{Zhao}, W.~{Lu}, M.~{Sheng}, Y.~{Chen}, J.~{Tang}, F.~R. {Yu}, and
  K.~{Wong}, ``{UAV}-assisted emergency networks in disasters,'' \emph{IEEE
  Wireless Commun.}, vol.~26, no.~1, pp. 45--51, Feb. 2019.

\bibitem{no_shadowing}
J.~G. {Andrews}, F.~{Baccelli}, and R.~K. {Ganti}, ``A tractable approach to
  coverage and rate in cellular networks,'' \emph{IEEE Trans. Commun.},
  vol.~59, no.~11, pp. 3122--3134, Nov. 2011.

\bibitem{NOMA_downlink_cellular}
K.~S. {Ali}, M.~{Haenggi}, H.~{ElSawy}, A.~{Chaaban}, and M.~{Alouini},
  ``Downlink non-orthogonal multiple access ({NOMA}) in poisson networks,''
  \emph{IEEE Trans. Commun.}, vol.~67, no.~2, pp. 1613--1628, Feb. 2019.

\bibitem{Table_of_integrals}
I.~S. Gradshteyn and I.~M. Ryzhik, \emph{Table of Integrals, Series and
  Products}.\hskip 1em plus 0.5em minus 0.4em\relax New York: Academic Press,
  6th ed, 2000.

\bibitem{Mathematics}
N.~Bourbaki, \emph{Elements of the History of Mathematics Paperback}.\hskip 1em
  plus 0.5em minus 0.4em\relax Springer Berlin Heidelberg, 2nd ed, 2008.

\end{thebibliography}
\end{spacing}

\end{document}